\documentclass[fleqn,usenatbib]{mnras}
\usepackage{CJKutf8}
\usepackage{newtxtext,newtxmath}
\usepackage[T1]{fontenc}

\DeclareRobustCommand{\VAN}[3]{#2}
\let\VANthebibliography\thebibliography
\def\thebibliography{\DeclareRobustCommand{\VAN}[3]{##3}\VANthebibliography}

\usepackage{graphicx}	
\usepackage{amsmath}	

\usepackage{threeparttable}
\usepackage{manyfoot}%
\usepackage{textcomp}%

\title[SNe Ia from CSST-UDF survey]{Cosmological Prediction of the CSST Ultra Deep Field Type Ia Supernova Photometric Survey}

\author[Wang et al.]{
Minglin Wang,$^{1,2}$
Yan Gong,$^{1,2,3}$\thanks{E-mail: gongyan@bao.ac.cn}
Furen Deng,$^{1,2}$
Haitao Miao,$^{1}$
Xuelei Chen,$^{1,2,4,5}$
and Hu Zhan$^{1,6}$
\\
$^{1}$National Astronomical Observatories, Chinese Academy of Sciences, Beĳing 100101, People’s Republic of China\\
$^{2}$University of Chinese Academy of Sciences, Beĳing 100049, People’s Republic of China \\
$^{3}$Science Center for China Space Station Telescope, National Astronomical Observatories, Chinese Academy of Sciences, \\20A Datun Road, Beĳing 100101, People’s Republic of China\\
$^{4}$Department of Physics, College of Sciences, Northeastern University, Shenyang 110819, China\\
$^{5}$Centre for High Energy Physics, Peking University, Beĳing 100871, China\\
$^{6}$Kavli Institute for Astronomy and Astrophysics, Peking University, Beijing 100871, China
}

\date{Accepted XXX. Received YYY; in original form ZZZ}

\pubyear{2023}
\usepackage{manyfoot}%
\begin{document}
\label{firstpage}
\pagerange{\pageref{firstpage}--\pageref{lastpage}}
\maketitle

\begin{abstract}

Type Ia supernova (SN Ia) as a standard candle is an ideal tool to measure cosmic distance and expansion history of the Universe. Here we investigate the SN Ia photometric measurement in the China Space Station Telescope Ultra Deep Field (CSST-UDF) survey, and study the constraint power on the cosmological parameters, such as the equation of state of dark energy. The CSST-UDF survey is expected to cover a 9 deg$^2$ sky area in two years with 250 s $\times$ 60 exposures for each band. The magnitude limit can reach $i\simeq26$ AB mag for 5$\sigma$ point source detection with a single exposure. We generate light curve mock data for SNe Ia and different types of core-collapse supernovae (CCSNe). {\tt SNCosmo} is chosen as the framework by utilizing the SALT3 model to simulate SN Ia data. After selecting high-quality data and fitting the light curves, we derive the light curve parameters and identify CCSNe as contamination, resulting in $\sim2200$ SNe with a $\sim7\%$ CCSN contamination rate. We adopt a calibration method similar to Chauvenet's criterion, and apply it to the distance modulus data to further reduce the contamination. We find that this method is effective and can suppress the contamination fraction to $\sim3.5\%$ with 2012 SNe Ia and 73 CCSNe. In the cosmological fitting stage, we did not distinguish between SNe Ia and CCSNe. We find that the constraint accuracies on $\Omega_{\rm M}$, $\Omega_{\Lambda}$ and $w$ are about two times better than the current SN surveys, and it could be further improved by a factor of $\sim$1.4 if including the baryon acoustic oscillation (BAO) data from the CSST spectroscopic wide-field galaxy survey.
\end{abstract}


\section{Introduction}

\begin{figure*}
  \centering
  \begin{minipage}[b]{0.49\textwidth}
        \includegraphics[width=1\linewidth]{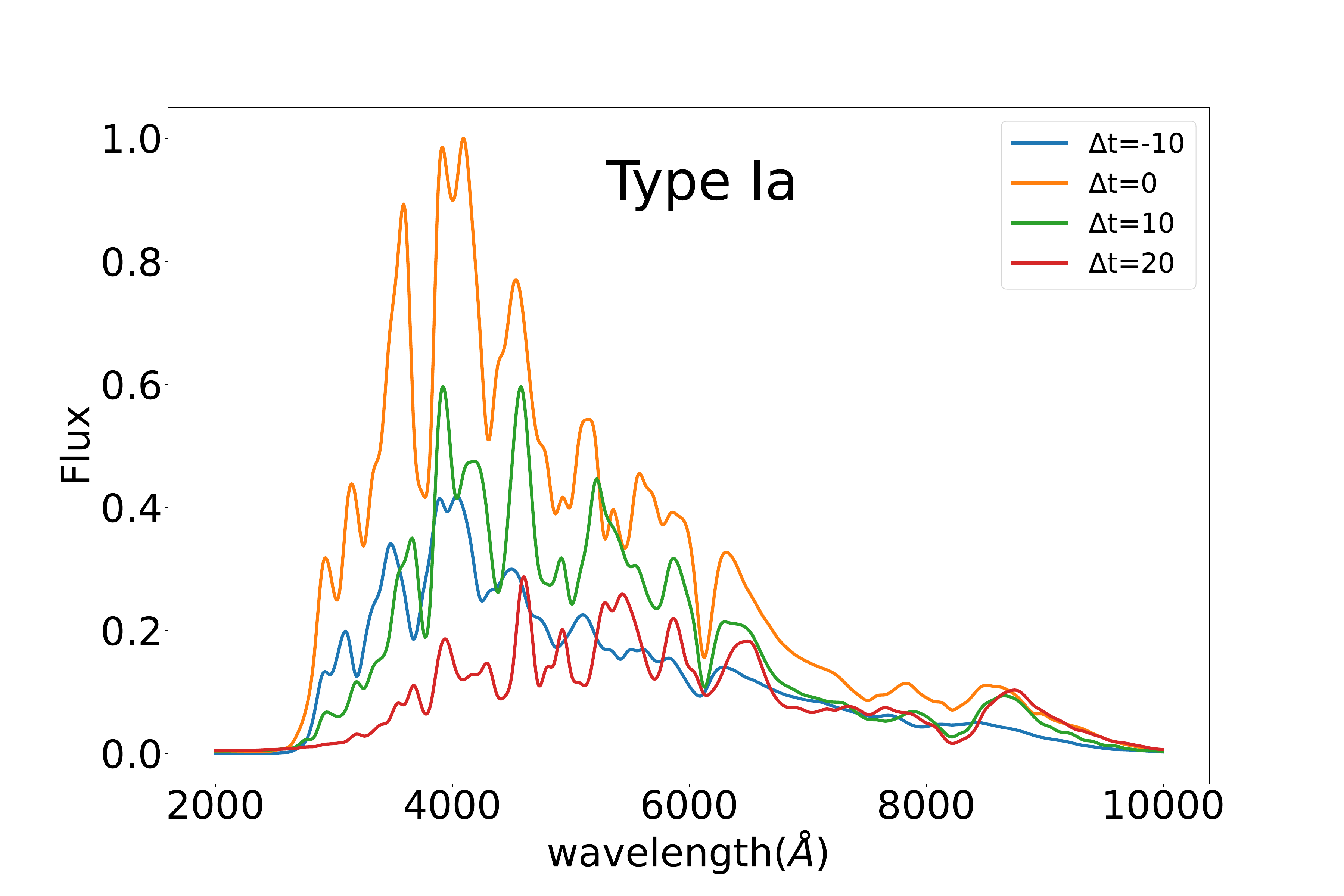}
    \centering
  \end{minipage}
  \hfill
  \begin{minipage}[b]{0.49\textwidth}
      \includegraphics[width=1\linewidth]{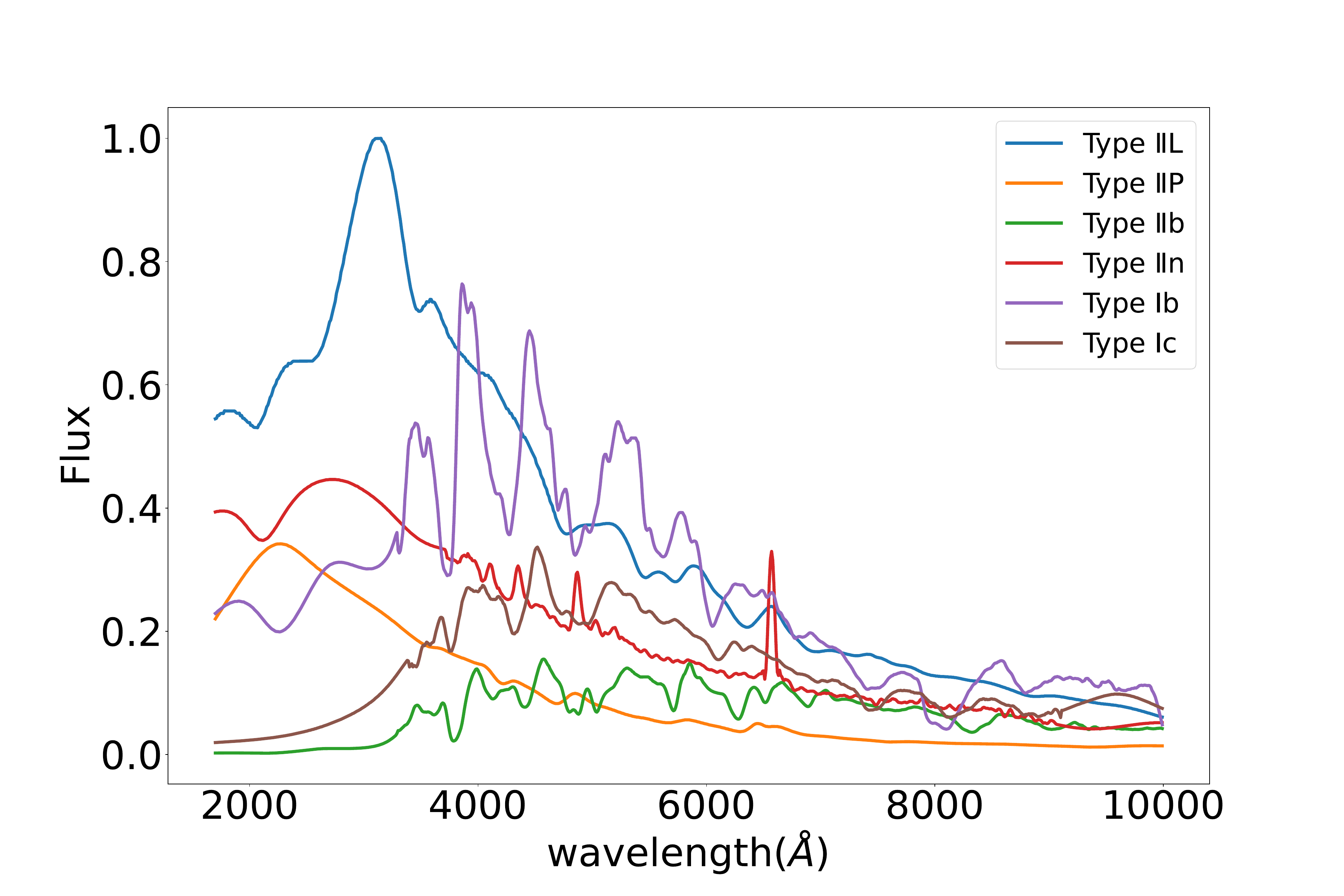}
  \end{minipage}
 \caption{ \label{fig:SNTEMPspec} The examples of the SED templets of SN Ia and CCSN adopted in this work. The $y$-axis shows the relative flux scales. {\it Left panel}: The SN Ia templates at various time points $\Delta \text{t}$ relative to the explosion time $t_0$ from the SALT3 model \citep{SALT3_Kenworthy2021}. {\it Right panel}: The randomly selected templates for the six CCSN types, i.e. SN IIL, SN IIP, SN IIb, SN IIn, SN Ib, and SN Ic, at the explosion time $t_0$ \citep{V19}.}
\end{figure*}

Type Ia supernova (SN Ia) plays a crucial role in the discovery of cosmic accelerating expansion and the study of the nature of dark energy \citep{Riess98,Perlmutter1999}. In the past two decades, significant advancements have been made in calibrating and expanding SN Ia samples across various redshift ranges. In recent years, many SN Ia surveys have been performed, such as Dark Energy Survey (DES, \citealt{DES_3YEARS_SN,DES5Year23,descollaboration2024dark}), Panoramic Survey Telescope and Rapid Response System (PanStarrs, \citealt{PANSTAR1_2014ApJ...795...44R}), Supernova Legacy Survey (SNLS, \citealt{SNLS2010}), Equation of State: SupErNovae trace Cosmic Expansion (ESSENCE, \citealt{ESSENCE_2007ApJ...666..674M}), and several Harvard-Smithsonian Center for Astrophysics (CFA) surveys \citep{CFA1_RIESS1999AJ....117..707R,CFA2_2006AJ....131..527J,CFA3_2009Hicken,CFA4_2012ApJS..200...12H}). Besides, there are deeper SN surveys conducted with the Hubble Space Telescope (HST), such as Great Observatories Origins Deep Survey
and Probing Acceleration Now with Supernova (GOODS+PANS, \citealt{goods2004,GOODS_Riess17...659...98R}), Supernova Cosmology Project (SCP, \citealt{SCP_suzuki_2012ApJ...746...85S}), Cosmic Assembly Near Infra-Red Deep Extragalactic Legacy Survey and Cluster Lensing And Supernova survey with Hubble (Candels+CLASH, \citealt{Candels_Riess2018ApJ...853..126R}), etc. Some combined SN~Ia samples are also recalibrated and generated as extensive datasets, such as Joint Light-curve Analysis (JLA, \citealt{JLA14}), Supernova H0 for the Equation of State (SH0ES, \citealt{SH0ES_2019ApJ...876...85R}), and Pantheon+ \citep{pahtheondataset}. However, most of the current SN Ia samples are based on ground-based telescopes, covering  relatively low redshift ranges. High-redshift SN~Ia samples at $z\gtrsim1$ are still extremely limited, which are important to accurately measure the cosmic expansion history and variation of the equation of state of dark energy.

In the next-generation ground- and space-based Stage IV surveys, such as the Vera C. Rubin Observatory (or LSST, \citet{LSST2019}), {\it Euclid} \citep{Euclid11,Euclid18}, Nancy Grace Roman Space Telescope (RST) (or WFIRST, \citet{WFIRST19}), and China Space Station Telescope (CSST \citealt{huzhan2011,Gong2019,HuZhan2021}), SN observation will be implemented as one of the main cosmological probes in these surveys. Thousands to millions of SNe~Ia are expected to be detected by these surveys in the next decade, and high-redshift SN Ia samples will be greatly expanded for precise cosmological study. However, since most of these SNe will be discovered photometrically, it would be quite challenging for spectroscopic follow-up measurements to accurately identify SNe Ia from the core-collapse supernovae (CCSNe) and determine their redshifts, given the huge amounts of SN data covering large redshift ranges. Therefore, it is necessary to explore the method of SN Ia confirmation and redshift estimation from the SN photometric data obtained by these future surveys, and assess the impacts on the relevant cosmological analysis.

Here we focus on the CSST SN photometric survey to investigate its capability of detecting and identifying SN Ia, and the constraint power on the dark energy equation of state and other cosmological parameters. The CSST is an upcoming Stage IV space-based telescope as a major science project of China Manned Space Program, which is expected to be launched in 2026 \citep{huzhan2011,Gong2019,HuZhan2021}. The diameter of its primary mirror is 2~m, and the field of view is 1.1 deg$^2$. It carries several astronomical instruments, including the survey camera, terahertz receiver, multichannel imager (MCI), integral field spectrograph (IFS), and coronagraph specialized in cool planet imaging. The CSST cosmological surveys performed by the survey camera plan to explore about 17,500 deg$^2$ and 400 deg$^2$ sky areas as the wide and deep field surveys, respectively, with multi-color photometric imaging and slitless spectroscopic observations in about ten years. Various important cosmological probes, such as weak and strong gravitational lensing, galaxy angular and redshift-space clustering, and galaxy clusters, will be carried out for studying the formation and evolution of cosmic large-scale structure and properties of dark energy and dark matter.

In addition, CSST also can perform an ultra-deep-field survey (CSST-UDF) covering $\sim$9 deg$^2$ sky area, which is planned to be located at high Galactic latitude. The single exposure time is 250~s with 60 exposures for each band in about two years. The magnitude limit in AB mag for point sources with 5$\sigma$ detection can reach $i\simeq28.2$ in the CSST-UDF survey \citep{CAOYE22}. For a single exposure, we can find that the magnitude limits can achieve 25.3, 25.7, 26.4, 26.1, 25.9, 25.4, 24.4 for the seven photometric bands covering $\sim$2500\AA\ to 10000\AA, i.e. $\it NUV$, $\it u$, $\it g$, $\it r$, $\it i$, $\it z$, and $\it y$. So CSST has great potential in SN Ia detection and photometric redshift (photo-$z$) estimation, especially at high redshifts, given the survey cadence and depth. Besides, as we discuss later, the host galaxy also can be detected with a certain probability by these bands, which can provide more information on the evaluation of redshift and dust extinction in the SN~Ia analysis. In \cite{LISHIYU}, they also explore the SN Ia detection and cosmological constraint by the CSST-UDF survey. Here we will consider the contamination from CCSNe, discuss the contamination reduction method, and adopt more realistic survey strategy.


In this work, we generate the light-curve mock data for both SN Ia and different types of CCSNe, based on the SN spectral energy distribution (SED) templates, the CSST instrumental design, and survey strategy of the CSST-UDF survey. We then use the SN Ia templates to fit the light-curve mock data at different CSST photometric bands, and try to distinguish and remove CCSNe according to the chi-square estimation. In order to further suppress the CCSN contamination, we also propose to use a method based on Chauvenet's criterion for identifying and eliminating CCSNe based on the derived data of the distance modulus. We investigate the effectiveness of this method, and perform the cosmological constraints on the equation of state of dark energy. The baryon acoustic oscillation (BAO) mock data from CSST spectroscopic survey are also included to reduce the degeneracies between different cosmological parameters.

The structure of this paper is as follow: Section~\ref{sec:MOCK} provides a detailed description of the CSST-UDF observation, SN templates, mock data generation, and light curve fitting process and result. In Section~\ref{sec:Distance Module}, we discuss the estimation of distance modulus, and the relevant contamination removal method. We present the cosmological constraint results for the non-flat $\Lambda$CDM model and flat $w$CDM model using various SN datasets and CSST BAO mock data in Section~\ref{sec:result}. Finally, we give summary and discussion in Section \ref{sec:summary}. We assume a  flat $\Lambda$CDM model as the fiducial cosmology with the dark matter energy density parameter $\Omega_\text{M}=0.3$ and the reduced Hubble constant $\text{h}=0.7$. 

\begin{table}
    \centering
    \caption{The natural generation rates (in units $\mathrm{yr}^{-1} \mathrm{Mpc}^{-3} 10^{-4} h_{70}^3$) of  SNe~Ia \citep{SN_RATE14} and CCSNe \citep{CCRATE——strolger15} we use in different redshift bins.}
    \begin{threeparttable}
    \begin{tabular}{|c|c|c|c|c|} \hline 
         \multicolumn{2}{|c|}{SN Ia}&  &  \multicolumn{2}{|c|}{CCSN}\\ \hline 
         Redshift&  Rate&  &  Redshift& Rate\\ \hline 
         0 - 0.5&  0.36&  &  0.1 - 0.5& 2.13\\ \hline 
         0.5 - 1&  0.51&  &  0.5 - 0.9& 3.86\\ \hline 
  1 - 1.5 & 0.64& & 0.9 -1.3&3.07\\ \hline 
 1.5 - 2& 0.72& & 1.3 - 1.7&3.25\\\hline
    \end{tabular}
    \end{threeparttable}
    \label{tab:snrate}
\end{table}

\begin{figure*}
  \centering
  \begin{minipage}[b]{0.48\textwidth}
        \includegraphics[width=1\linewidth]{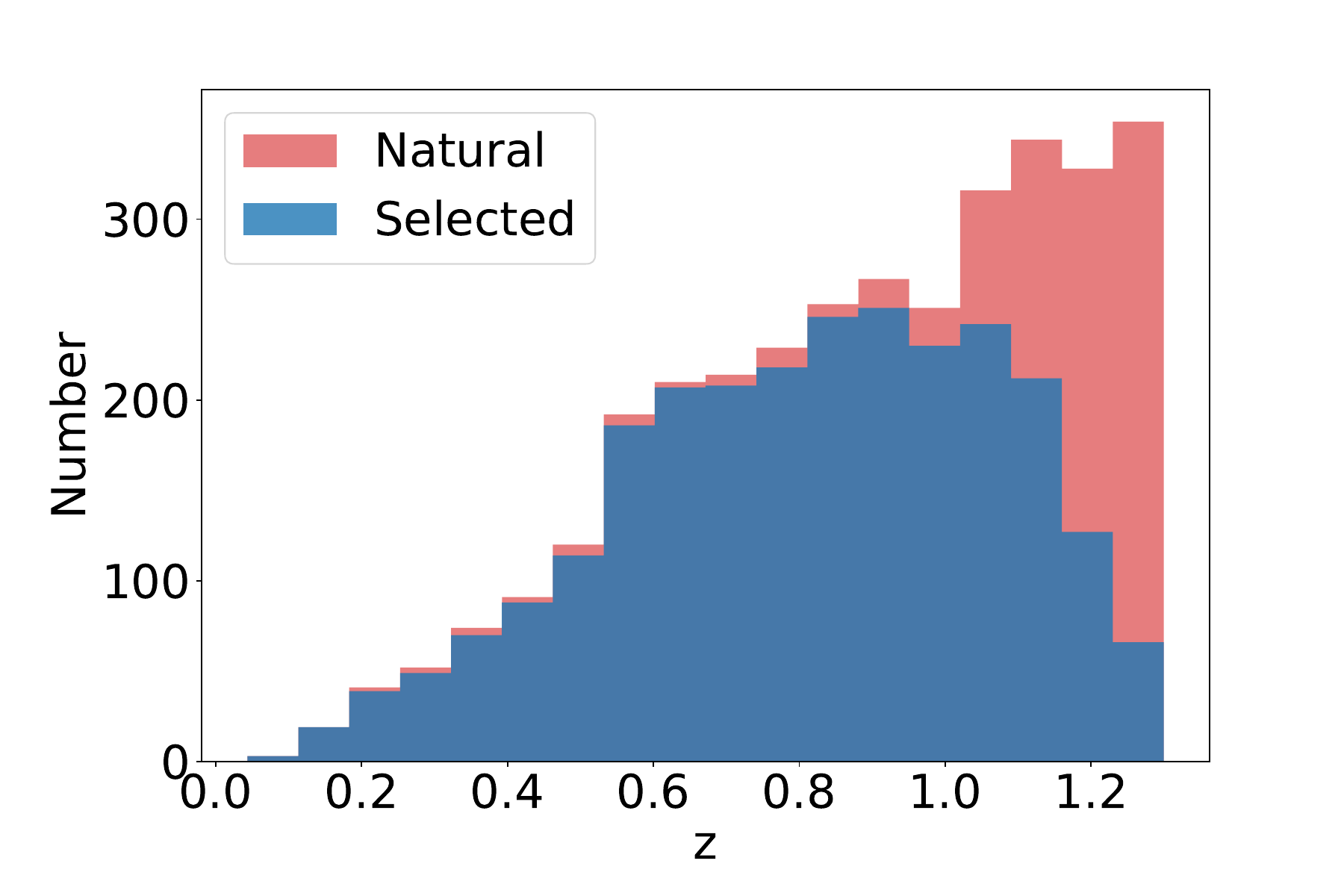}
    \centering
  \end{minipage}
  \hfill
  \begin{minipage}[b]{0.48\textwidth}
      \includegraphics[width=1\linewidth]{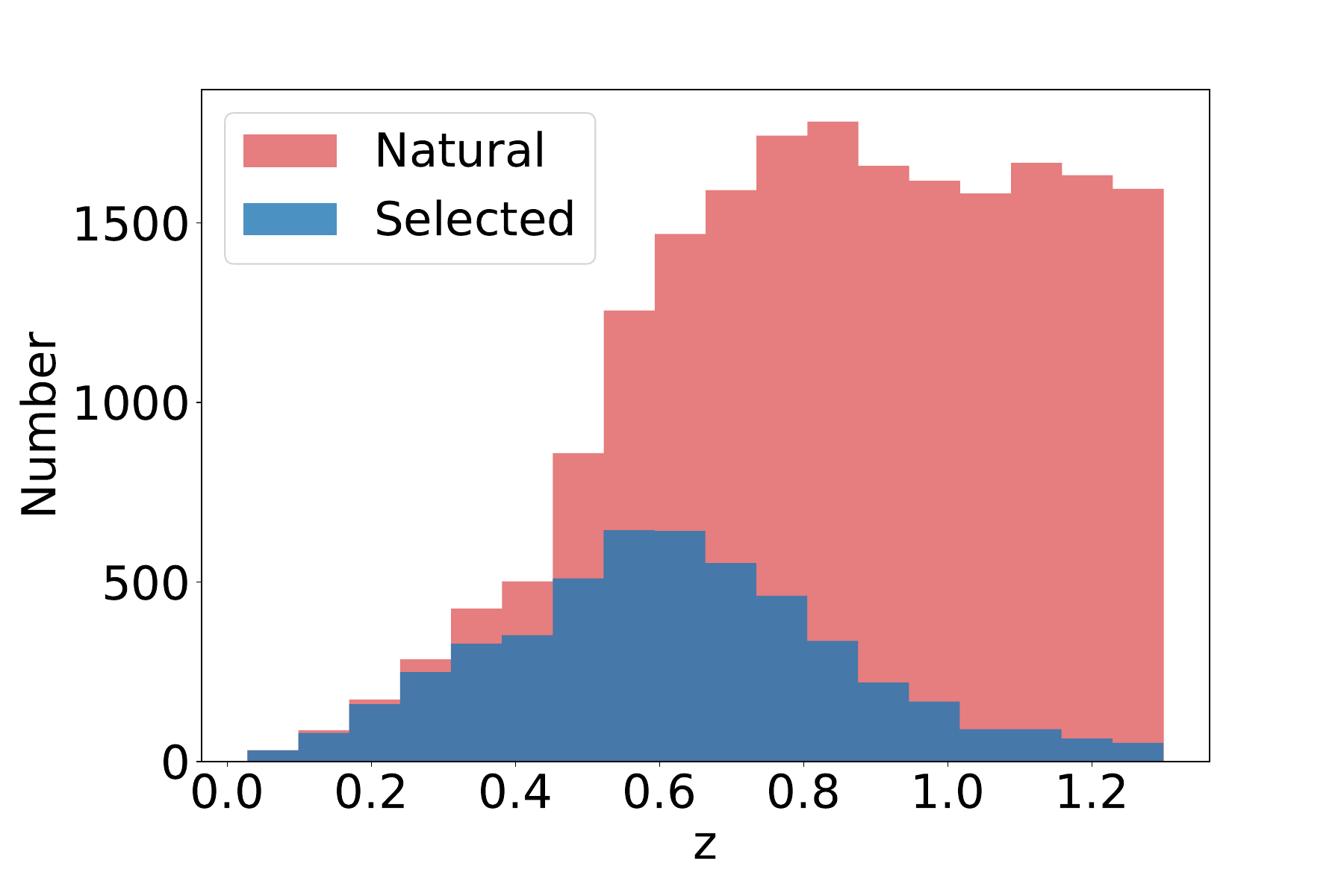}
  \end{minipage}
 \caption{ \label{fig:image4} The redshift distributions of SNe Ia (left panel) and CCSNe (right panel) in 9 deg$^2$ in two years. The red and blue histograms show the natural generation SN sample and selected sample for the CSST-UDF survey, respectively.}
\end{figure*}

\section{Light curve mock data} \label{sec:MOCK}

We employ {\tt SNCosmo} \citep{SNCOSMO} as a tool to generate and fit the mock light curve data.  {\tt SNCosmo} is a supernova cosmology simulation Python package, which is similar to {\tt SNANA} \citep{SNANA}. It includes functions such as light curve generation and parameter fitting, and is easy to modify and customize. Based on the SN SED templates, we first generate natural SN populations for both SN Ia and CCSN, and then derive a mock light curve SN catalog with flux and error for the CSST photometric bands, considering the CSST-UDF survey strategy. After applying selection criteria, we can obtain the final mock light curve data.

\subsection{SN template} \label{subsubsec: salt3} 

  \begin{figure}
    \includegraphics[width=1\linewidth]{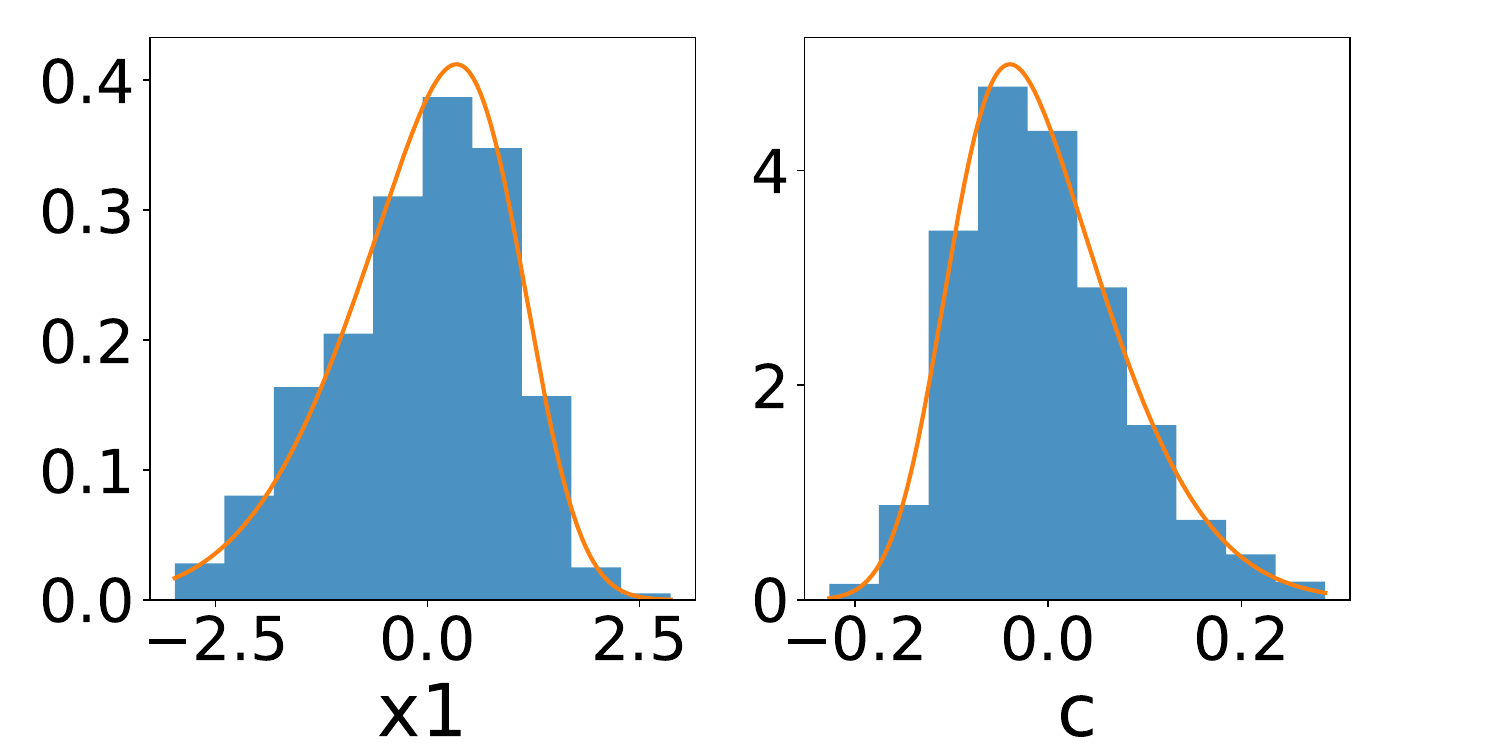}
    \caption{The distributions of $x_1$ and $c$ parameters derived from the Pantheon+ \citep{pahtheondataset}. The orange curves show the fitted skewed normal distributions used to randomly sample the values of these two parameters.}
    \label{fig:x1andc}
\end{figure}

Various SN models currently exist, including but not limited to MLCS2k2 \citep{MLCS2K2}, SALT2 \citep{SALT2_2007}, and SUGAR \citep{SUGAR20}. Among these, the SALT3 model \citep{SALT3_Kenworthy2021} represents the third generation of the SALT model, and has been widely applied in SN surveys, such as Pantheon \citep{pahtheondataset}, DESSN \citep{DES_3YEARS_SN}, and SNLS \citep{SNLS2010}.

SALT3 is an upgraded version of SALT2, which extends spectral coverage and reduces systematic errors, while maintaining the linear transformability of model parameters from SALT2 to SALT3. Here we choose to use the SALT3 model in this study, and the spectral flux or SED template of a given SN Ia in the SALT3 model can be expressed as  \citep{SALT3_Kenworthy2021}
\begin{equation} \label{eq:Fp}
    F(p, \lambda)=  x_0\left[M_0\left(p, \lambda ; \boldsymbol{m}_{\mathbf{0}}\right)+x_1 M_1\left(p, \lambda ; \boldsymbol{m}_{\mathbf{1}}\right)\right] 
 \cdot \exp[c \cdot C L(\lambda ; \boldsymbol{c l})].
\end{equation}
Here, $p=t/(1+z)$ is the SN rest-frame phase, $\lambda$ is the wavelength, $x_0$ describes the amplitude of the light curve, $x_1 $ characterizes the time-dependent variation, and $c$ is the color parameter that is independent of time. \(M_0(p, \lambda;\boldsymbol{ m_0})\) represents the SED of a SN Ia, while \(M_1(p, \lambda; \boldsymbol{m_1})\) reflects the time-dependent variation. \(CL(\lambda;\boldsymbol{ cl})\) is a single color law accounting for the effect of $c$ on the light curve by including the effects of intrinsic color variation and dust extinction of host galaxy, which can be described by polynomials. $\boldsymbol{m_0}$, $\boldsymbol{m_1}$, $\boldsymbol{cl}$ are model parameter vectors which can be  obtained from the model calibrating process. We show the examples of the SN Ia SED templates we use at different time relative to the explosion day in the left panel of Figure~\ref{fig:SNTEMPspec}.

On the other hand, for the CCSNe, we adopt the templates proposed by \cite{V19} (hereafter V19). This library contains 67 time-series SED templates from photometric and spectroscopic observations, and includes various CCSN types, such as SN IIP, SN IIL, SN IIn, SN IIb, SN Ib, and SN Ic. The right panel of Figure \ref{fig:SNTEMPspec} displays the SED templates of six CCSN types as examples randomly selected from the V19 library.

\subsection{Light curve generation} \label{subsec:lc gen}

After obtaining the SN templates, we can generate the light curve mock data based on the CSST instrumental design and survey strategy. For SNe Ia, the SALT3 model is used for light curve generation, which contains five parameters, i.e. $z$, $t_0$, $x_0$, $x_1$, and $c$. To determine the SN Ia redshift distribution, we adopt the natural SN Ia generation rate given by \cite{SN_RATE14}, which is derived from the Cosmic Assembly Near-infrared Deep Extragalactic Legacy Survey (CANDELS) and other surveys. We find that there are about $\sim$3300 SNe Ia can be generated in 9 deg$^2$ at $0<z<1.3$ in two years. The details of the SN Ia natural generation rate and redshift distribution can be found in Table~\ref{tab:snrate} and the left panel of Figure~\ref{fig:image4} (red histogram). For the explosion time $t_0$, we assume an uniformly random explosion rate in two years for these SNe~Ia. The values of $x_1$ and $c$ are extracted from the distributions given by \citet{pahtheondataset} as shown in Figure~\ref{fig:x1andc}. We have fitted the distributions using skewed normal distributions (orange curves), and randomly select values of $x_1$ and $c$ from these distributions for each SN Ia. The amplitude $x_0$ can be automatically generated by the SALT3 model, after $x_1$, $c$, and the absolute magnitude are determined. The absolute magnitude for a SN Ia can be expressed by $M_{\text{abs}}=M_0-(\alpha x_1-\beta c)$. Here we assume $M_0$ follows a normal distribution with a mean value equals to -19.25 with the intrinsic scatter $\sigma=0.1$, and $\alpha=0.16$, $\beta=3$  \citep{MABS-19.253RIESS,descollaboration2024dark}.

\begin{table}
\centering
\caption{\label{tab:luminosity-function-probability}The relative fractions and LFs for various types of CCSNe used in this work  \citep{V19}. The $R$-band LFs follow the normal distribution, and the mean values and $\sigma$ have been shown here.}
\begin{threeparttable}
\begin{tabular}{|l|c|c|}
\hline
\text{SN Type} & \text{Fraction}& \text{LF (mean, $\sigma$)}\\
\hline
SN II-L& 7.9\%& -18.28 (0.45)\\
SN II-P& 57\%& -16.67 (1.08)\\
SN II-b& 10.9\%& -16.69 (1.99)\\
SN II-n & 4.7\%& -17.66 (1.08)\\
SN Ic & 11.9\%& -17.44 (0.66)\\
SN Ib & 7.5\%& -18.26 (0.15)\\
\hline
\end{tabular}
\end{threeparttable}
\end{table}

For CCSN, we need to determine the natural generation rate, luminosity function (LF), and relative fractions for different types. We utilize the CCSN natural generation rates at different redshift bins provided by \cite{CCRATE——strolger15}, as shown in Table~\ref{tab:snrate} and the right panel of Figure~\ref{fig:image4}. This generation rate is derived from CANDELS and Cluster
Lensing And Supernova survey with Hubble (CLASH), and we totally obtain $\sim20,000$ CCSNe in 9 deg$^2$ in two years. Regarding to the CCSN LFs, we adopt the simulation results given in \citet{V19}, utilizing the data provided by \cite{J17}. The CCSN LFs are assumed to be normal distributions, and the mean values and errors of LFs and relative fractions for different CCSN types can be found in Table~\ref{tab:luminosity-function-probability}. Note that we actually adopt high CCSN LFs in \citet{V19}, which could lead to relatively high contamination on SN Ia sample. This means the CCSN contamination can be smaller if using low CCSN LFs, and the assumption we make here is conservative.

Since the extinction of host galaxy for SNe Ia has already been considered in the SALT3 model by the color law functuon, we only consider the extinction of Milky Way $E^\text{MW}_\text{(B-V)}$ with reddening law $R_V$=3.1 \citep{f99}. Given that the CSST-UDF should locate at high Galactic latitudes, we assume that $E^\text{MW}_\text{(B-V)}$ is around 0.01 \citep{vincenzi2024desSystematic}, and it can be randomly generated from a normal distribution with mean value 0.01 and scatter $\sigma$ = 0.001 for each SN. Note that, for simplicity, we do not consider the extinction of host galaxy for CCSNe, which undoubtedly will enhance the effect of the CCSN contamination.
     \begin{figure}
          \centering
          \includegraphics[width=1\linewidth]{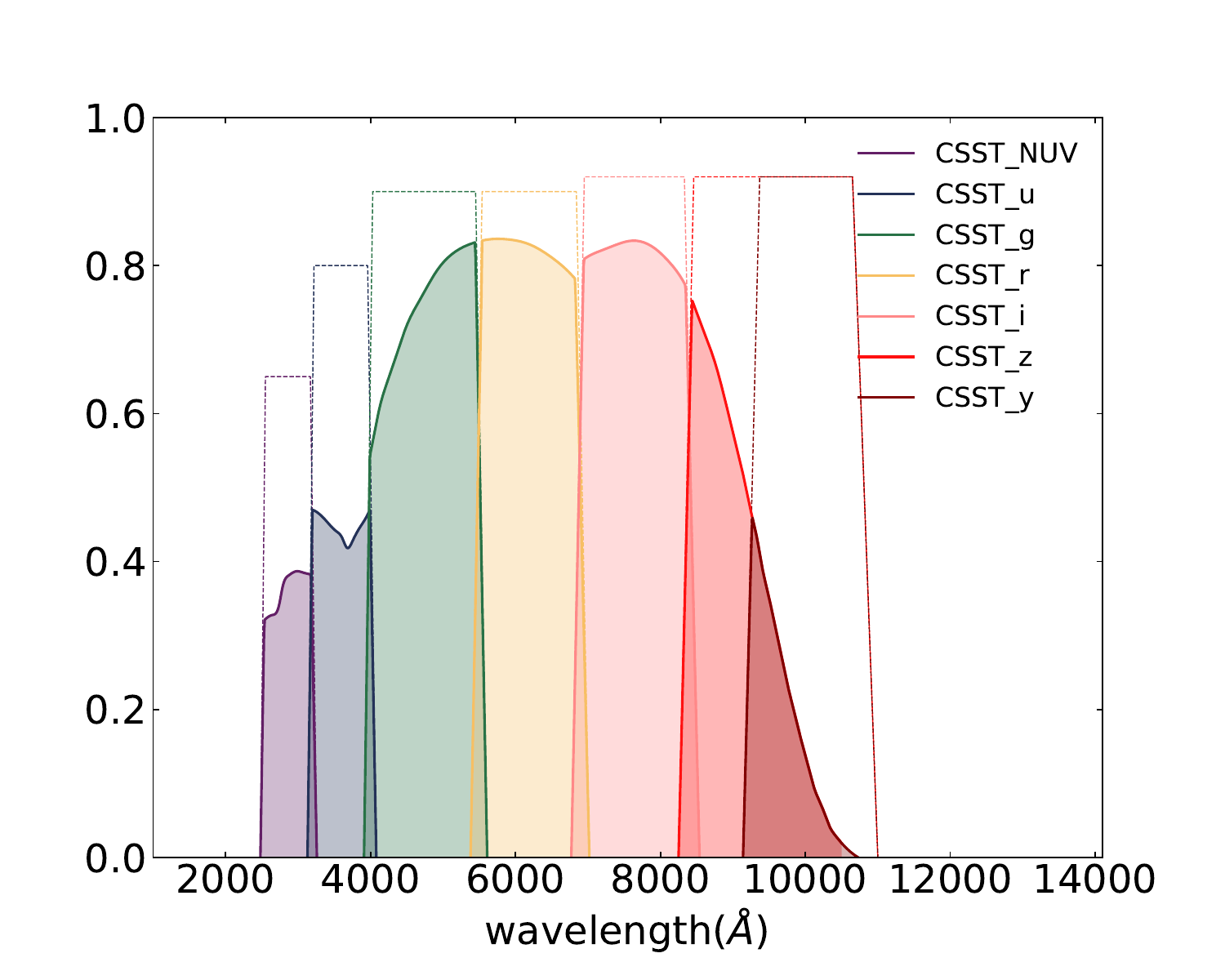}
          \caption{The filter transmission curves for the seven CSST photometric bands. The dotted curves denote the intrinsic  transmissions, and the solid ones are the total transmissions including the detector quantum efficiency.}
          \label{fig:TRANS}
      \end{figure}

Then the flux of a SN observed in a certain band at time $t$ can be estimated by
\begin{equation}
f_b(t)=\int T_b\left(\lambda\right) F\left(t, \lambda \right) \lambda d \lambda,
\end{equation}
Where $F(t, \lambda)$ is the spectral flux for SN Ia (given by Equation~\ref{eq:Fp}) or CCSN, as shown in Figure~\ref{fig:SNTEMPspec}. $T_b$ is the transmission function in a specific wavelength band. The filter transmission curve for each CSST band has been shown in Figure~\ref{fig:TRANS}.
To estimate the flux error, we can calculate the signal-to-noise ratio (SNR), which can be calculated by \citep{caoye18,ACSBOOK} :
     \begin{equation}
\text{SNR}=\frac{C t}{\sqrt{C t+N_{\text{pix}}\left(B_{\text{sky}}+B_{{\text{det }}}\right) t+N_{\text{pix}} N_{\text {read}} R^2}}.
\label{eq:snr}
\end{equation}
Here $C$ is the signal from the source in $e^-\,s^{-1}$, $B_{\text{det}} \leq0.02e^-/\text{s}/\text{pix}$ is the dark current, $N_\text{read}=1$ is the number of readouts, $R =5e^-/\text{pix}$ is the readout noise, and $N_\text{pix}$ is the number of pixels occupied by the target on the CCD, which can be estimated according to the pixel size and PSF size. For simplicity, we assume $N_\text{pix}$=16, which is a conservative estimate.  $B_\text{{sky}}$ is the sky background noise. For the CSST seven photometric bands, $B_\text{{sky}}=0.003$, 0.018, 0.156, 0.200, 0.207, 0.123 and 0.036 $e^-/\text{s}/\text{pix}$, respectively.

For the SN survey cadence, since the CSST-UDF survey strategy has not been fully determined, we simply assume an uniform cadence for all CSST bands with 60 exposures in two years for a given field, which means a 12-day cadence. This cadence is close to the best cadence proposed by \citet{LISHIYU}, which can detect the largest number of SNe. The effect of the cadence will be studied in our future work.

The SN light curve mock data can be generated as discussed above, and then we need to select the high-quality SN Ia data for cosmological analysis. Our selection criteria are as follows:

\begin{enumerate}
  \item One photometric measurement with SNR>5 before the peak brightness and two after the peak brightness in one band.
    \item One photometric measurement with SNR>5 before the peak brightness and one after the peak brightness in another band.
    \item At least 1 photometric measurement with SNR>10.
    \item At least 6 photometric measurements with SNR>5 in all bands.
\end{enumerate}

After applying the selection criteria, the numbers of SNe Ia and CCSNe are reduced to $\sim$2600 and $\sim$5000, respectively. The redshift distributions of the selected SNe Ia and CCSNe are shown in blue histograms of Figure~\ref{fig:image4}. We notice that the selection criteria we adopt are stringent compared to some current photometric SN surveys, e.g. DES \citep{DES_3YEARS_SN}. Although it may lose some SN samples, as we will discuss later, this stringent selection can provide high-quality data, reduce the CCSN contamination, and improve the accuracy of the cosmological constraint.

\begin{figure*}
  \centering
  \begin{minipage}[b]{0.48\textwidth}
    \centering
    \includegraphics[width=\textwidth]{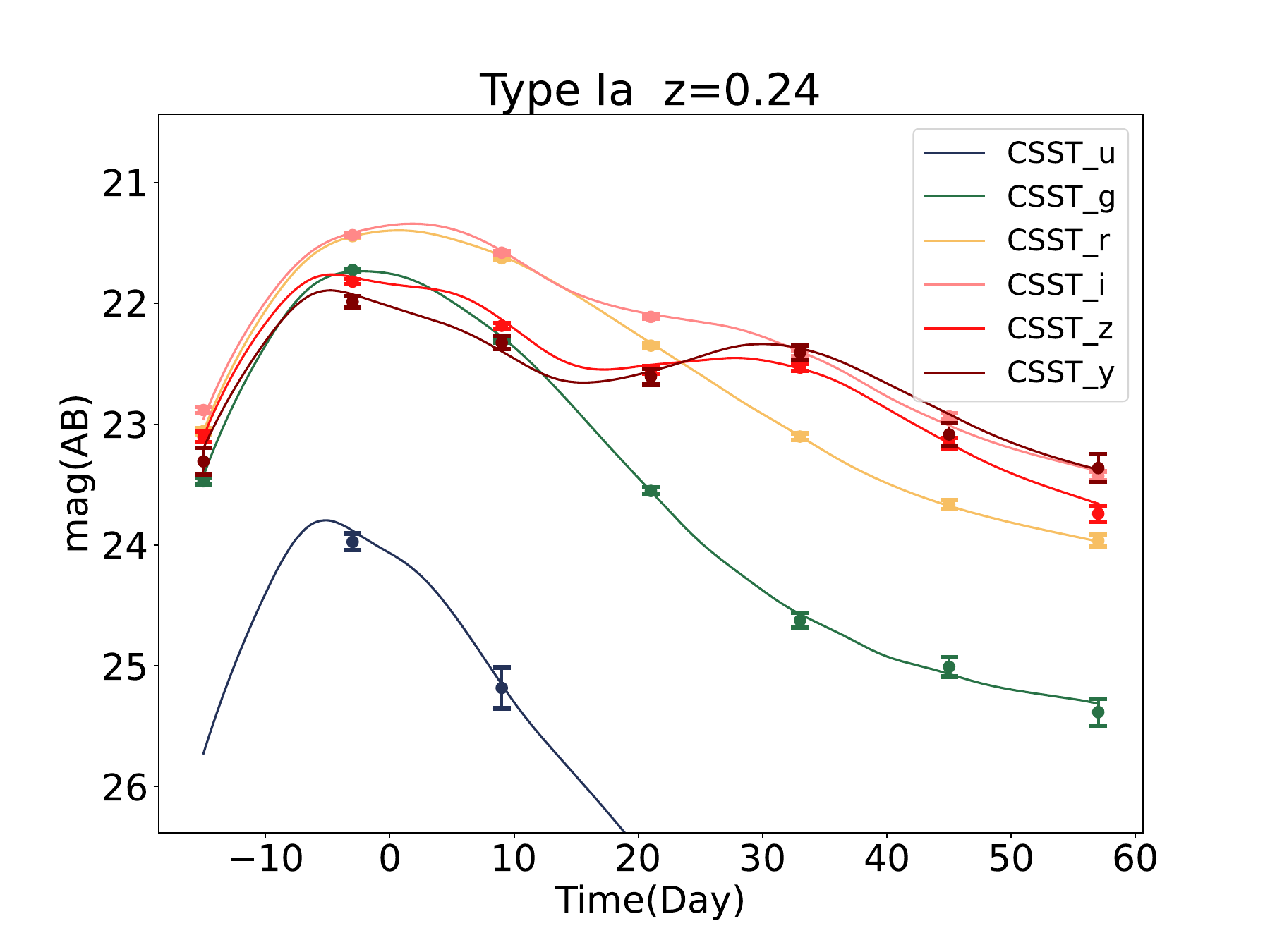}
  \end{minipage}
  \hfill
  \begin{minipage}[b]{0.48\textwidth}
    \includegraphics[width=\textwidth]{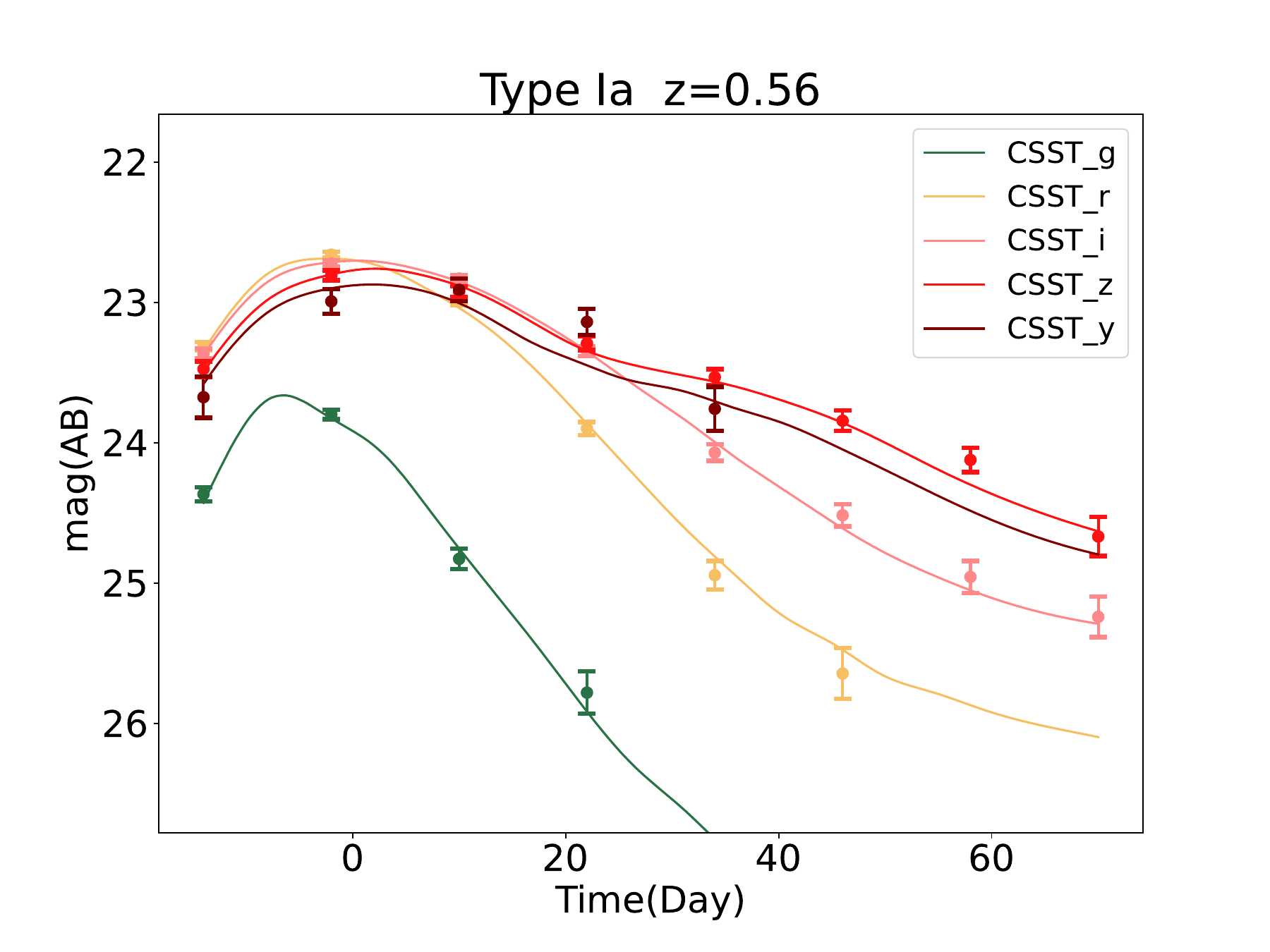}
  \end{minipage}
\vspace{0.5\baselineskip}
  \begin{minipage}[b]{0.48\textwidth}  

      \centering
      \includegraphics[width=1\linewidth]{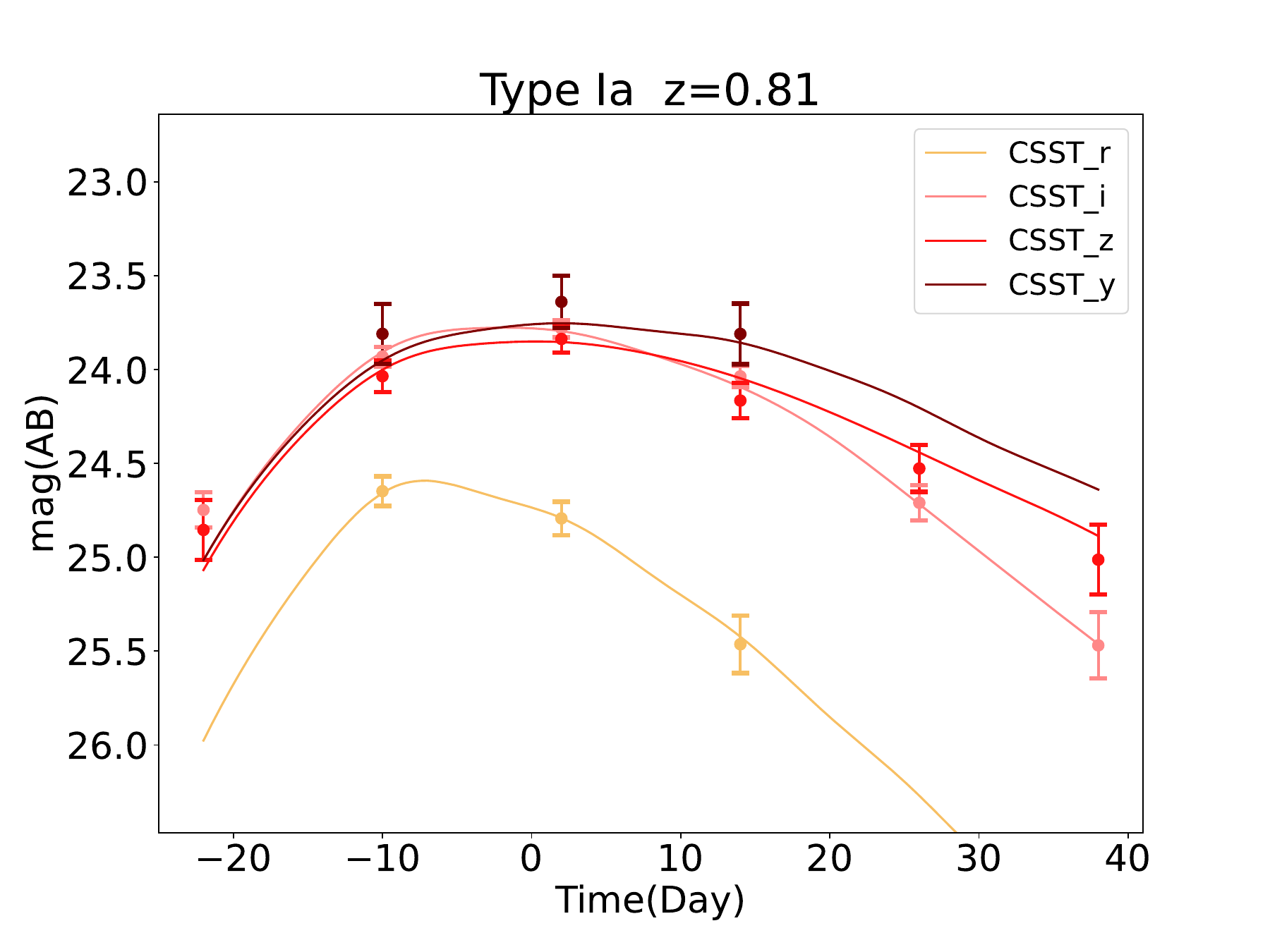}
\end{minipage}
  \hfill
  \begin{minipage}[b]{0.48\textwidth}
        \includegraphics[width=1\linewidth]{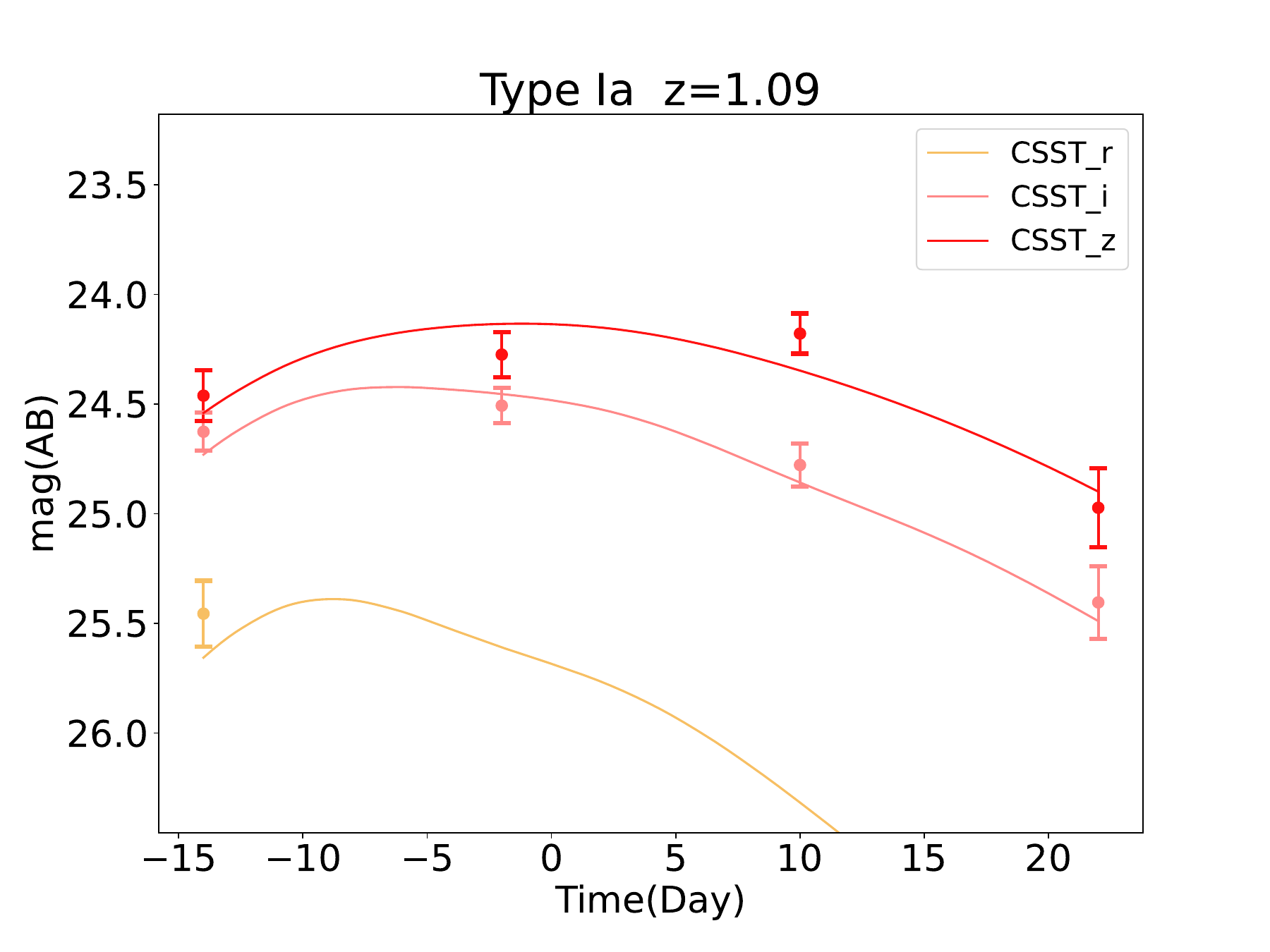}
        \end{minipage}
 \caption{  \label{fig:Iasnlightcurve}The four mock light curves of SNe Ia at various redshifts from $z=0.2$ to 1.1 for different photometric bands in the CSST-UDF survey. The solid lines are the theoretical light curves derived from the fiducial model.}
\end{figure*}

\begin{figure*}
    \hfill
    \centering
    \vspace{0.5\baselineskip}
  \begin{minipage}[b]{0.48\textwidth}
       \includegraphics[width=1\linewidth]{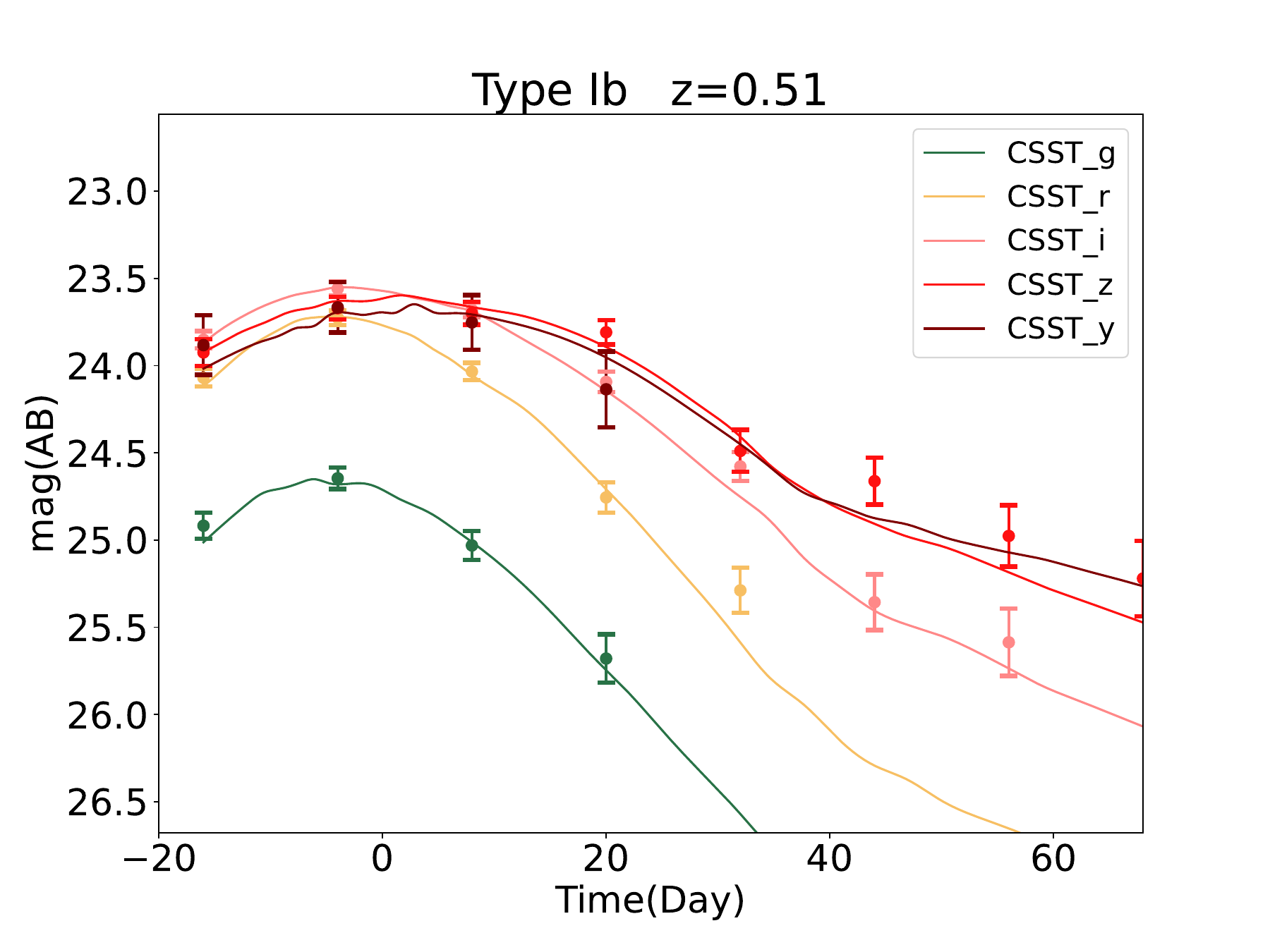}
     \end{minipage}    
    \hfill
    \vspace{0.5\baselineskip}
  \begin{minipage}[b]{0.48\textwidth}
      \includegraphics[width=1\linewidth]{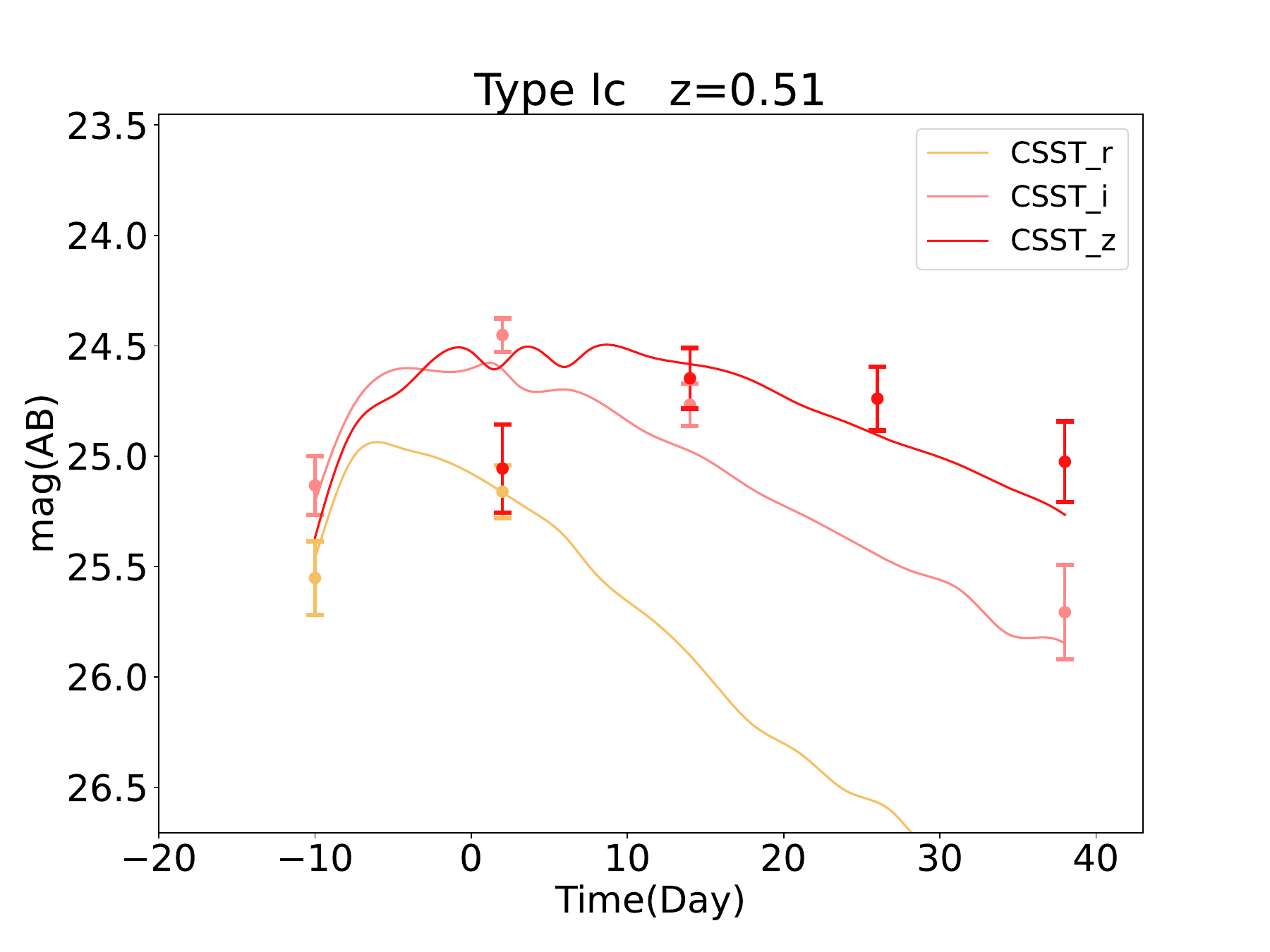}
    \end{minipage}
   \hfill
    \centering
    \vspace{0.5\baselineskip}
  \begin{minipage}[b]{0.48\textwidth}
       \includegraphics[width=1\linewidth]{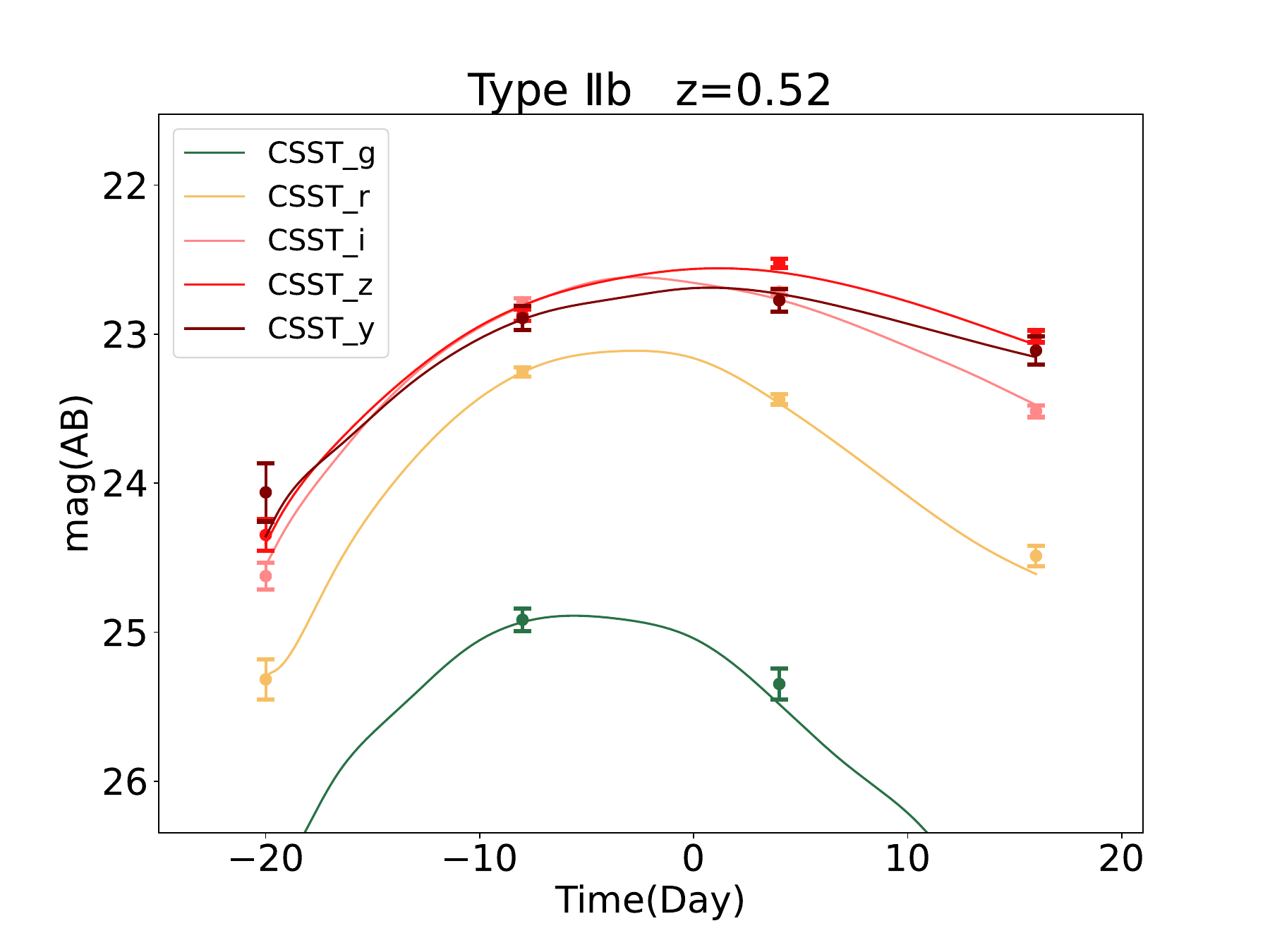}
     \end{minipage}    
    \hfill
    \vspace{0.5\baselineskip}
  \begin{minipage}[b]{0.48\textwidth}
      \includegraphics[width=1\linewidth]{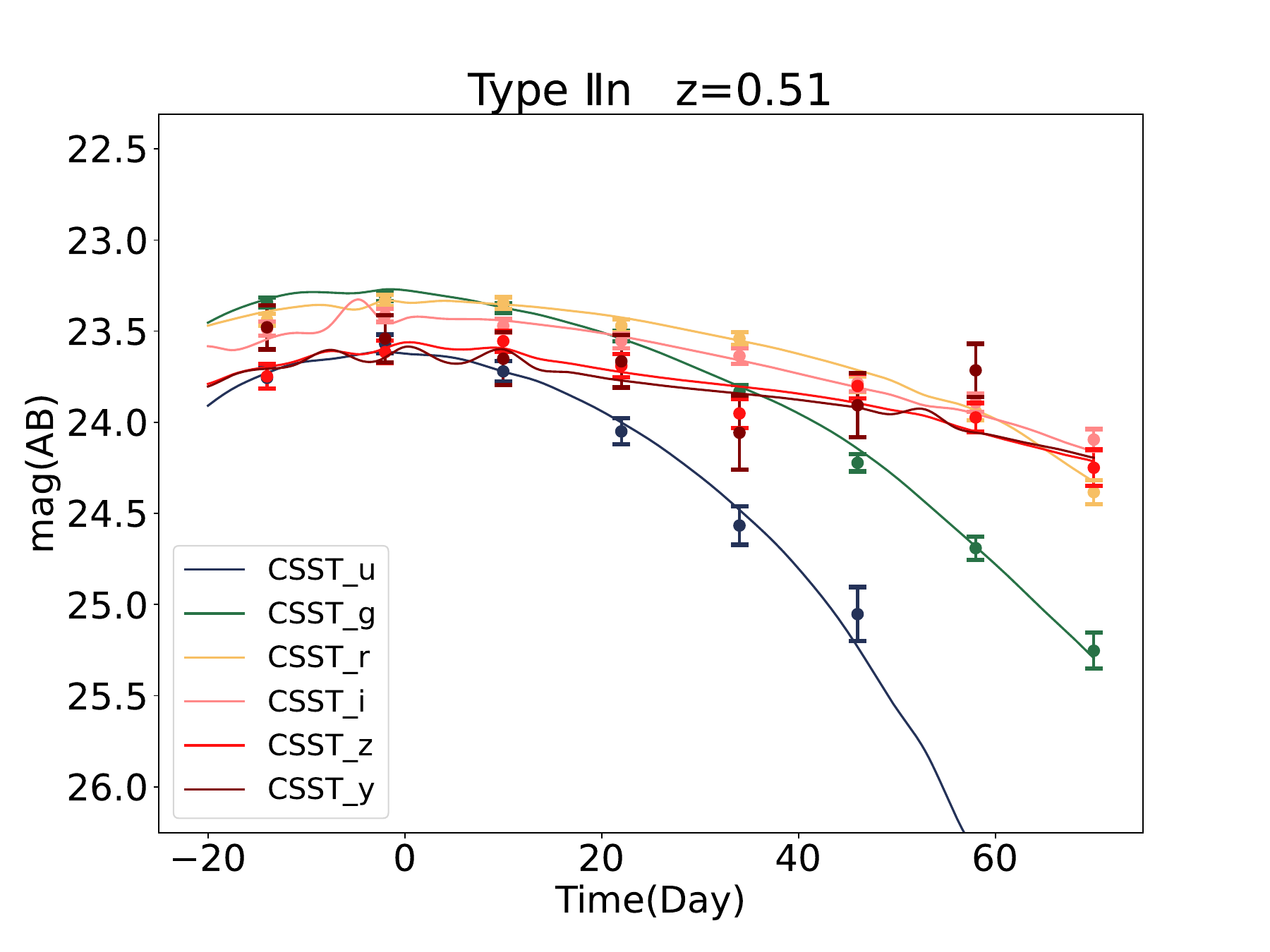}
    \end{minipage}
   \hfill
    \centering
    \vspace{0.5\baselineskip}
  \begin{minipage}[b]{0.48\textwidth}
       \includegraphics[width=1\linewidth]{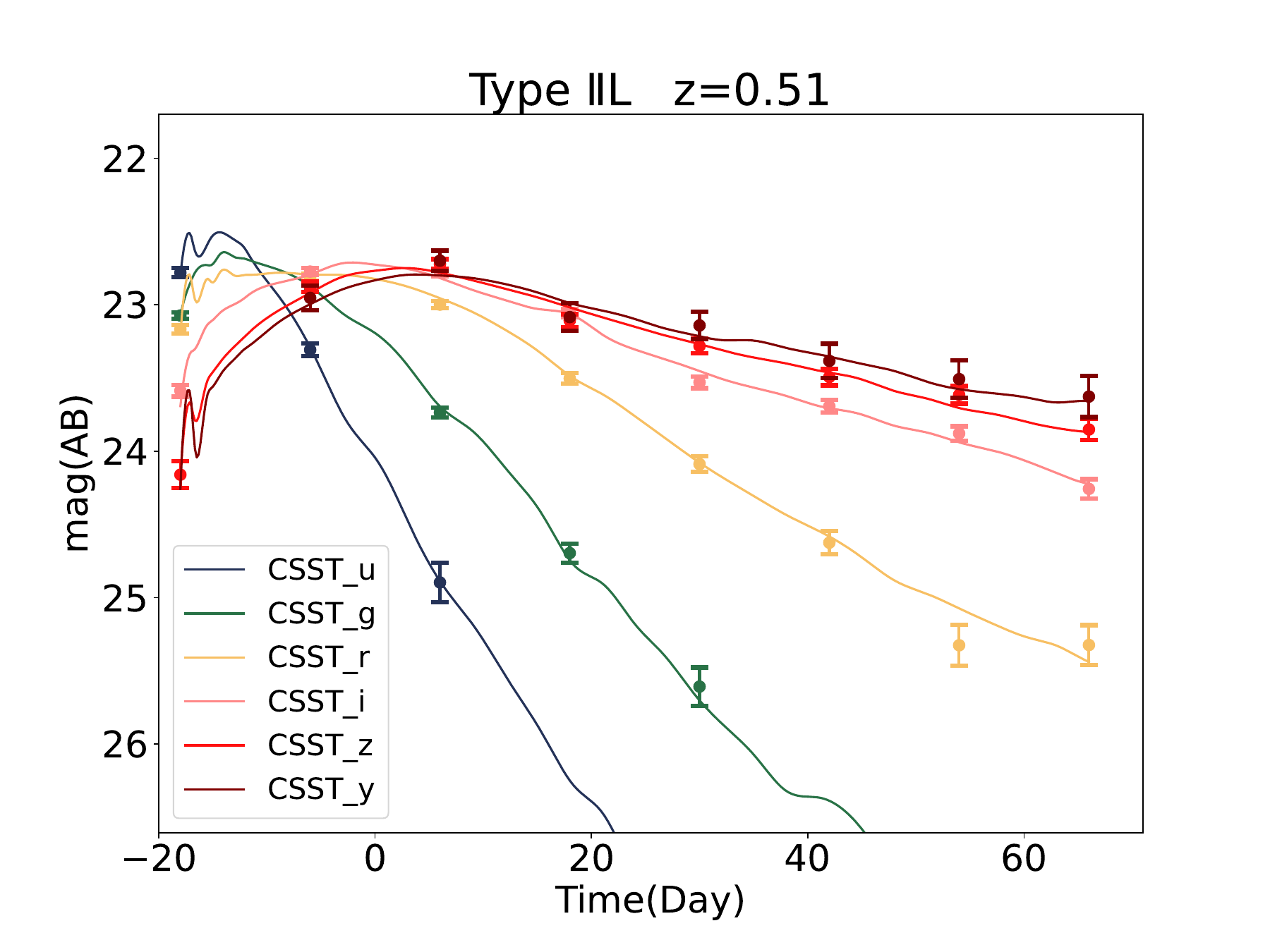}
     \end{minipage}    
    \hfill
    \vspace{0.5\baselineskip}
  \begin{minipage}[b]{0.48\textwidth}
      \includegraphics[width=1\linewidth]{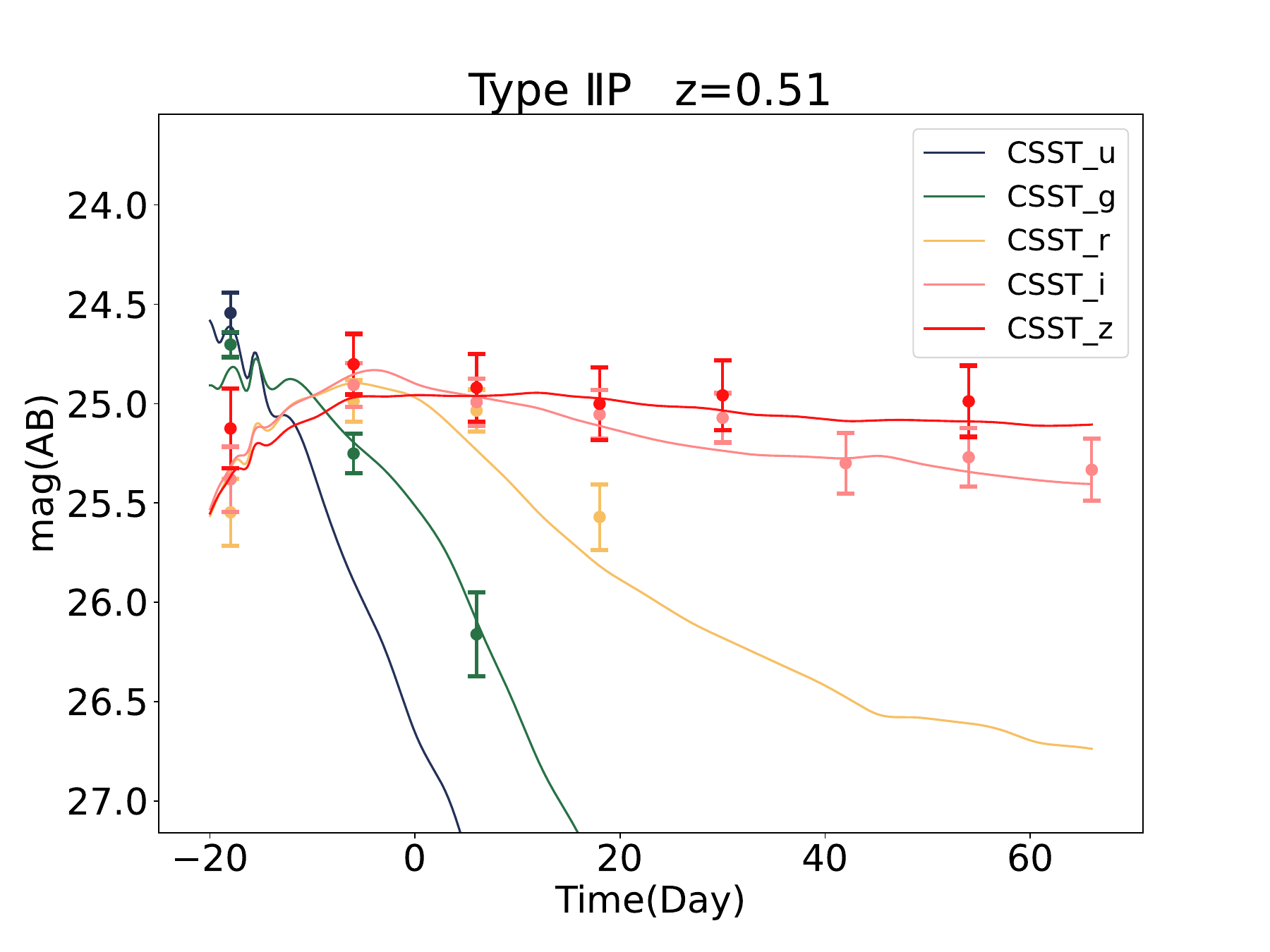}
    \end{minipage}
 \caption{\label{fig:ccsnlightcurve}The examples of the mock light curves for the six types of CCSNe considered in this work at $z\sim0.5$. The solid lines are the theoretical light curves derived from the fiducial model.}
\end{figure*}

In Figure~\ref{fig:Iasnlightcurve} and Figure~\ref{fig:ccsnlightcurve}, as examples, we show the light curve mock data of SNe Ia and CCSNe, respectively, after applying the selection criteria, measured in the CSST-UDF for different photometric bands. The theoretical curves from the fiducial model are also shown for comparison. Note that the data points of the light curves are not exactly overlapped with the theoretical curves from the fiducial model, since the data points are randomly generated from the normal distributions based on the theoretical  flux and errors.

 \subsection{Light curve fitting result }\label{subsec:light curves}
  \begin{table}
      \centering
\caption{The predicted detection rate of host galaxy photo-$z$ with 24 AB mag as the apparent magnitude limit in different redshift ranges \citep{DESHOSTGALAXY}. We assume that the rate is about 40\% at $z\in(1.2, 1.3)$.}
\label{tab:host rate}
\begin{threeparttable}
      \begin{tabular}{|c|c|} \hline 
           Redshift range  &  Detection rate\\ \hline 
           0 - 0.1&  98\%\\ \hline 
           0.2 - 0.3&  97\%\\ \hline 
           0.3 - 0.4&  94\%\\ \hline 
           0.4 - 0.5&  92\%\\ \hline 
           0.5 - 0.6&  89\%\\ \hline 
           0.6 - 0.7&  85\%\\ \hline 
 0.7 - 0.8& 82\%\\ \hline 
 0.8 - 0.9& 78\%\\ \hline 
 0.9 - 1.0& 74\%\\ \hline 
 1.0 - 1.1& 70\%\\ \hline 
 1.1 - 1.2& 67\%\\ \hline
 1.2 - 1.3&40\%\\\hline
      \end{tabular}
      \end{threeparttable}
  \end{table}

Based on the light curve mock data from the CSST-UDF survey, next we can fit them to explore the constraint power on the parameters of the SN Ia light curve and remove the CCSN contamination. The Markov Chain Monte Carlo (MCMC) method is adopted for fitting the light curve parameters using the built-in MCMC program {\it sncosmo.nest\_lc()} in {\tt SNCosmo}. We have five parameters here, and flat priors (or parameter ranges) are assumed in the fitting process. We have $z\in(0,1.3)$, $x_1\in(-3.5,3.5)$, $c\in(-0.35,0.35)$, and the ranges of $t_0$ and $x_0$ can be obtained automatically by $\tt SNCosmo$ based on the flux data points \citep{SNCOSMO}. 

We also consider the photo-$z$ measurements of host galaxies to improve the SN light curve fitting. A $\pm5\%$ photo-$z$ uncertainty of host galaxy is assumed and used as a prior when fitting the light curve data. The detection rate or fraction of SN host galaxies with the photo-$z$ measurement is given by \citet{DESHOSTGALAXY} and shown in Table \ref{tab:host rate}. We randomly select the SN samples based on this detection rate of host galaxy. Note that this host galaxy detection rate is for the apparent magnitude limit of 24 AB mag, which corresponds to about one-exposure depth (250 s exposure time) in the CSST-UDF survey. This is obviously a conservative assumption, since the magnitude limit can reach $\sim$27 AB mag for galaxies using the complete CSST-UDF survey with 60 exposures. Higher magnitude limit can result in higher host galaxy detection rate, that could improve the SN Ia light curve fitting result.
The rate of host galaxy 
In order to improve the efficiency, we first make use of the least squares method to fit the data, and find the best-fit values of the light curve parameters. These best-fit values can be set as the initial values for the free paramters in the MCMC process. The MCMC program given in {\tt SNCosmo} is employing the nest algorithm, utilizing a single-ellipsoidal method. A maximum iteration of 50,000 is set to prevent the MCMC from stalling, which is especially useful when fitting the CCSN data. 

This fitting process also can help to remove the CCSN contamination. We propose three selection criteria to remove the contamination after obtaining the fitting results of the light curve parameters, and the SN samples will be rejected if they satisfy:
\begin{enumerate}
    \item The reduced chi-square $\chi^2_{\text{reduce}} >3 $.
    \item The best-fit values of $x_1$ or $c$ excess the ranges $x_1\in(-3,3)$ and $c\in(-0.3,0.3)$, required by the SALT3 model.
    \item One of the Errors of $t_\text{0}$, $x_\text{1}$ and $c$ excess the ranges: $t_\text{0err}>2$, $x_\text{1err}>1$ and $c_\text{err}>0.1$.

  \end{enumerate}

\begin{figure}
    \centering
    \includegraphics[width=1\linewidth]{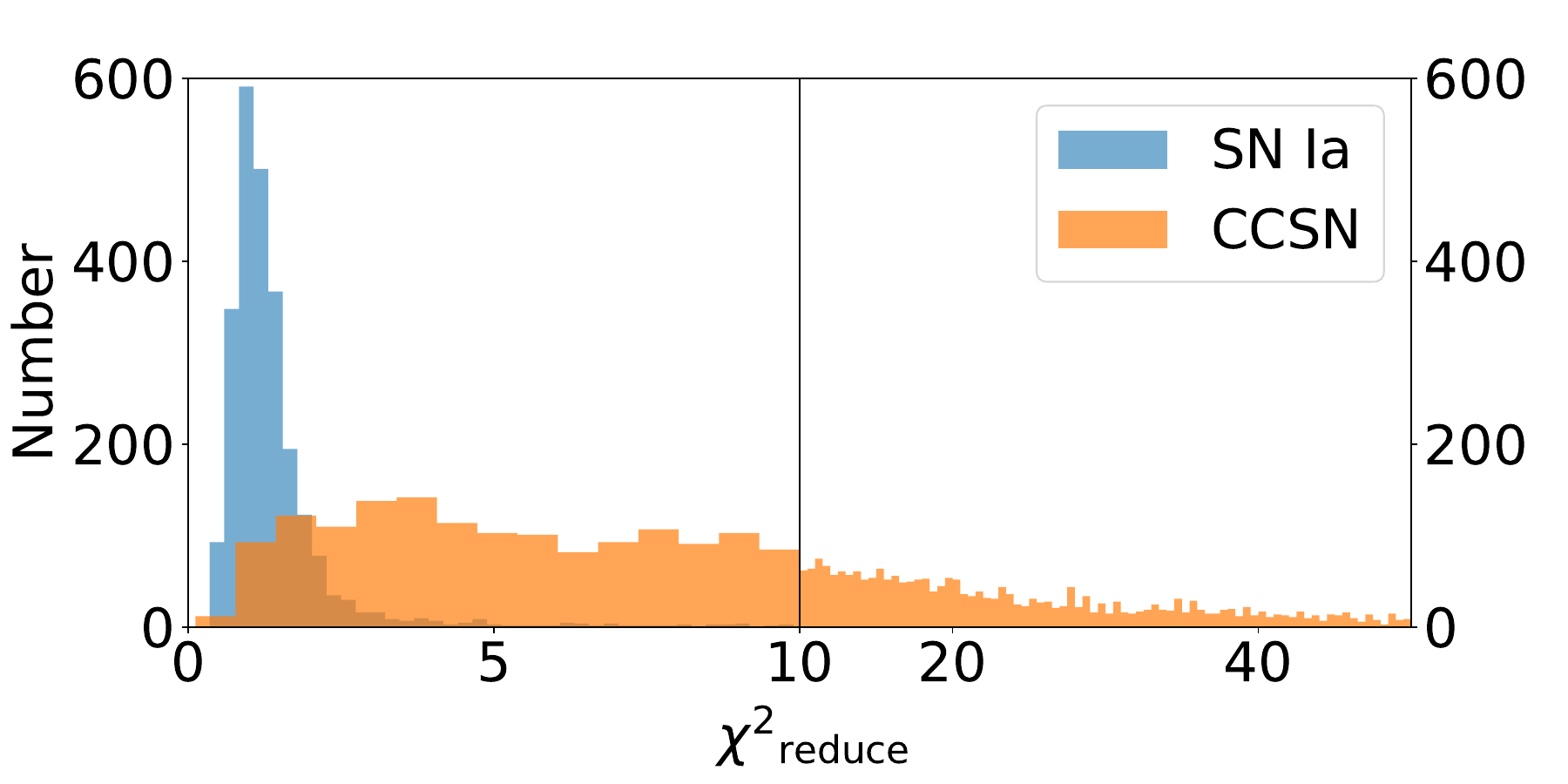}
    \caption{The $\chi^2_{\text{{reduce}}}$ distributions for SNe Ia (blue) and CCSNe (orange) in the light curve fitting process. The SN samples with $\chi^2_{\text{{reduce}}}>3$ have been removed to reduce the CCSN contamination.}
    \label{fig:rsq_template_fitting}
\end{figure}


\begin{figure}
        \centering
        \includegraphics[width=1\linewidth]{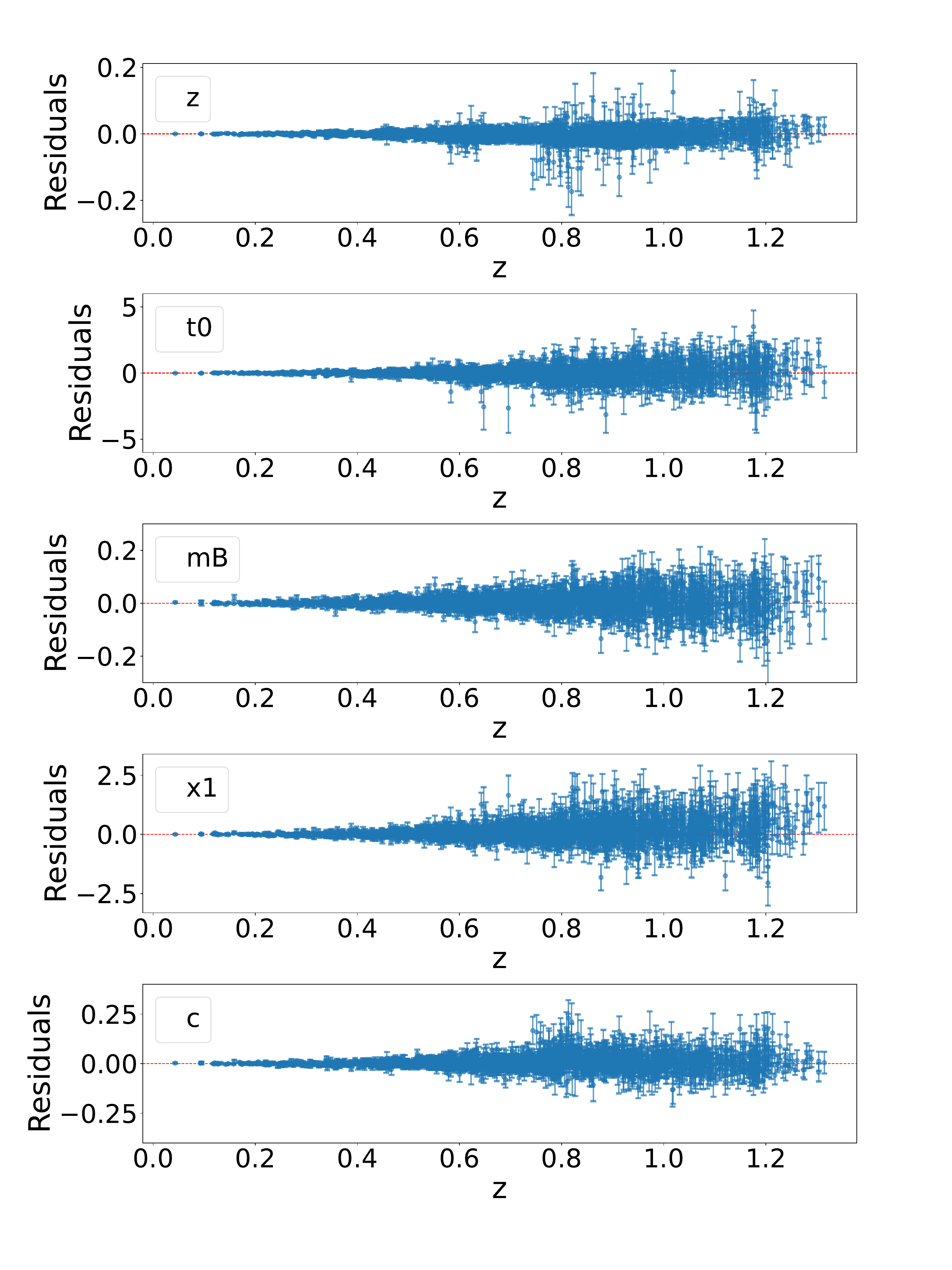}
        \caption{The residuals of the best-fit values of the five light curves parameters relative to the fiducial values for the pure SN Ia sample. The error bars show 1$\sigma$ uncertainties of the parameters derived from the MCMC chains.}
        \label{fig:parameters}
\end{figure}

Using these three selection criteria, we find that 555 SNe Ia and 4854 CCSNe can be removed, and the remaining SNe Ia and CCSNe are 2020 and 155, respectively. In Figure~\ref{fig:rsq_template_fitting}, we show the $\chi^2_{\text{{reduce}}}$ distribution for both SNe Ia and CCSNe in the fitting process. We can see that most of SNe Ia has small reduced chi-square with $\chi^2_{\text{{reduce}}}<2$, while most of CCSNe have large $\chi^2_{\text{{reduce}}}$ that many of them can be even greater than 10. This means that the $\chi^2_{\text{{reduce}}}$ selection criterion should be effective to remove the CCSN contamination in our analysis. In Figure~\ref{fig:parameters}, the residuals of the best-fit values of the five light curve parameters relative to the fiducial values for the pure SN Ia sample have been shown. Note that we show the $B$-band apparent magnitude $m_{B}$ instead of $x_0$ here, and the relation between these two parameters is $m_B=10.5-2.5{\rm log}_{10}(x_0)$ \citep{SALT3_Kenworthy2021}. We can find that basically accurate fitting results can be obtained for these parameters at $z<0.7$, and the results become worse at higher redshifts, especially at $z>0.8$. 

As we show above, the light curve fitting process seems effective to derive the SN Ia light curve parameters, including $z$, $t_0$, $x_0$ or $m_B$, $x_1$, and $c$, and remove CCSNe (with $\sim$7.1\% contamination left) in the SN sample. However, we notice that the contamination should be further suppressed to get accurate cosmological constraint results, given the data quality of the CSST-UDF survey. This could be probably achieved by analyzing the distance modulus as we discuss below.


\section{Distance Modulus Calibration}\label{sec:Distance Module}


Using the fitting results of the light curve parameters, i.e. $m_B$, $x_1$ and $c$, we can calculate the distance modulus for the $i$th SN in the sample, which is given by
\begin{equation}
    \mu_i = m_{B,i} + \alpha x_{1,i} -\beta c_i- M_0.
    \label{eq:tripp2}
\end{equation}
Here $\alpha$, $\beta$, and $M_0$ are nuisance parameters, and we set them as free parameters in the fitting process. The error of distance modulus consists of three parts, i.e.
\begin{equation}
    \sigma_{i}^2=\sigma_{\mathrm{int}}^2+\sigma^2_{\mu,z} +{\sigma^2_{\mu,i}}.
    \end{equation}
Here $\sigma_{\mathrm{int}}=0.1$ is the intrinsic dispersion of the SN Ia absolute magnitude \citep{SALT3_Kenworthy2021,vincenzi2024desSystematic}, $\sigma_{\mu, z}$ is the error from the photo-$z$ uncertainty $\sigma_z$, which has the relation as $\sigma_{\mu,z}=\frac{5}{\ln (10)} \frac{1+z}{z(1+z / 2)} \sigma_z$ \citep{SDSS2_zerror}, and $\sigma^2_{\mu,i}$ is the error accounting for the uncertainties of $m_B$, $x_1$ and $c$ for each SN, which can be derived by their covariance matix obtained in the light curve fitting process.

On the other hand, the theoretical distance modulus is related to the luminosity distance $d_{\rm L}$, which is determined by the cosmological model, and we have
\begin{equation}
\mu_{\rm th}(z)=5 \log _{10} d_{\rm L}(z)+25.
\label{eq:MUmodel}
\end{equation}
Here $d_{\rm L}$ is given by 
\begin{equation}
    d_{\rm L}=(1+z) \int_0^z \frac{c\,{\rm d} z^{\prime}}{H\left(z^{\prime}\right)},
    \label{eq:DL}
\end{equation}
where  \(H(z)\) is the Hubble parameter that takes different forms under various cosmological models. In this work, it can be expressed as
\begin{equation}
    H(z)=H_\text{0} \left[\Omega_{\mathrm{M}}(1+z)^3+\Omega_{\rm DE}(1+z)^{3(1+w)}+\Omega_{\rm k}(1+z)^2\right]^{1 / 2}.
\label{eq:ezzz}
\end{equation}
Hence the cosmological parameter vector $\boldsymbol{\theta}_{\rm c}=(\Omega_{\rm M}, \Omega_{\rm DE}, w, H_0)$, where $\Omega_{\rm M}$ and $\Omega_{\rm DE}$ are the energy density parameters of dark matter and dark energy, $\Omega_{\rm k}=1-\Omega_{\rm M}-\Omega_{\rm DE}$ is the curvature energy density parameter, $w$ is the dark energy equation of state, and $H_0$ is the Hubble constant. In this study, we mainly explore two cosmological models, i.e. the flat $w$CDM and nonflat $\Lambda $CDM model. In the flat $w$CDM model, since $\Omega_{k}= 0$ and $\Omega_{\rm DE}=1-\Omega_{\mathrm{M}}$, we focus on $w$, $\Omega_{\mathrm{M}}$, $H_\text{0}$. In the nonflat $\Lambda$CDM model, we have $w$=-1, and the free parameters are $\Omega_{\rm DE}=\Omega_{\Lambda}$, $\Omega_{\mathrm{M}}$, and $H_\text{0}$. Besides the cosmological parameters, we also have $\alpha$, $\beta$ and $M_0$ as the nuisance parameters in the model.

\begin{figure}
    \centering
    \includegraphics[width=1\linewidth]{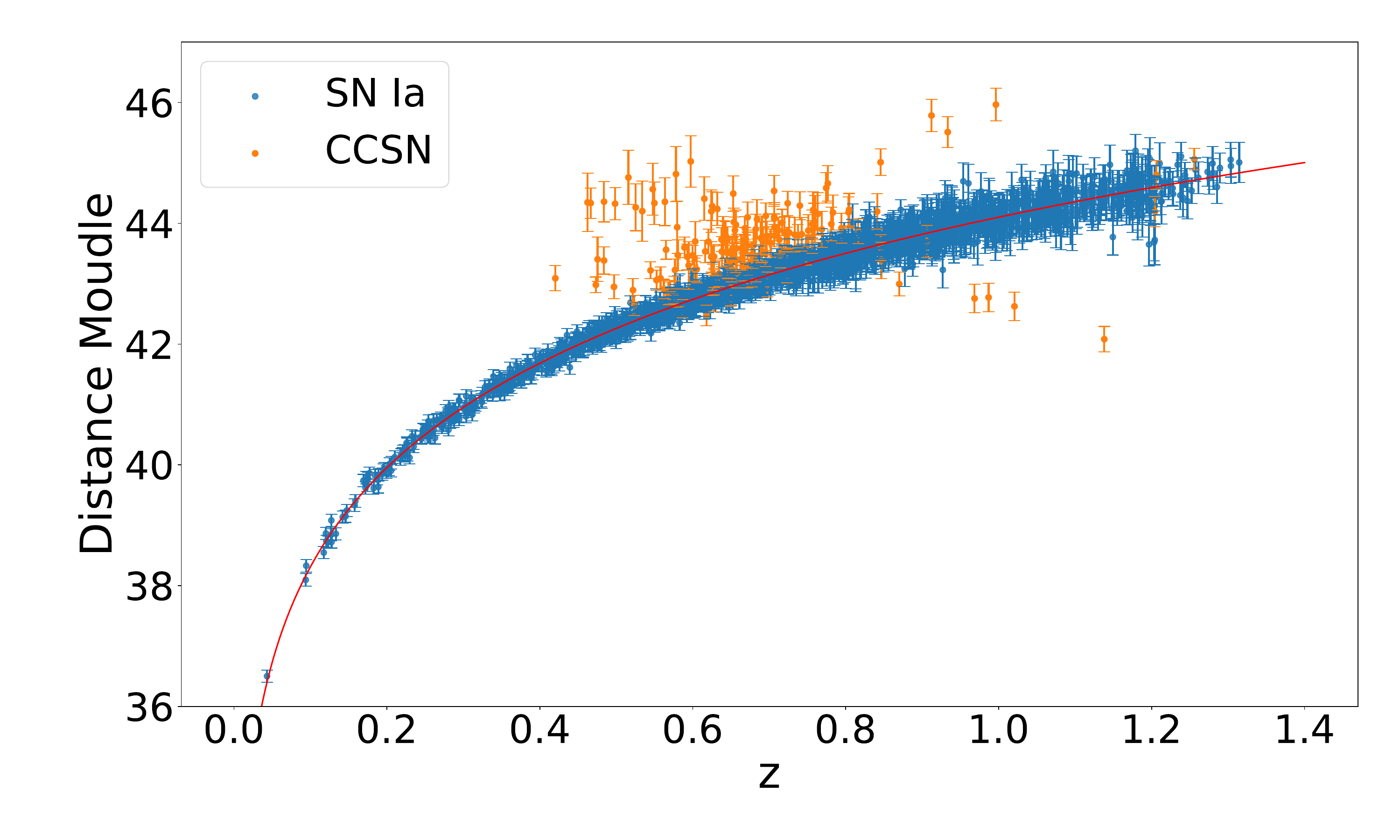}
    \caption{The Hubble diagram vs. the input redshifts for the SN samples after the selection of the light curve fitting process. The blue and orange data points are for SNe Ia and CCSNe, respectively. For comparison, the theoretical distance modulus from the fiducial flat $\Lambda$CDM model adopted in this work is also shown in red curve.}
    \label{fig:hdall}
\end{figure}

In Figure~\ref{fig:hdall}, we show the Hubble diagram for the input redshifts. To show it clearly, the values of $\alpha$, $\beta$ and $M_0$ are fixed to be the fiducial values as shown in Section~\ref{subsec:lc gen} when generating the data points, and the theoretical curve is obtained by assuming the fiducial flat $\Lambda$CDM model. We can find that most of the SN Ia data are overlapped with or closely around the the theoretical curve, which could correctly indicate the underlying cosmology. On the other hand, the rest of the CCSNe after the selection of the light curve fitting process are mainly at $0.4<z<0.8$ (CCSNe at low redshifts have good data quality and are easily rejected, and CCSNe at $z>0.8$ are hardly detected in the CSST-UDF survey), and most of them have large deviation from the SN Ia data or the theoretical curve as outliers. This provides an effective way to further remove the CCSNe from the SN dataset. 

Here we propose to use a method based on Chauvenet's criterion to remove the CCSN contamination by fitting the distance modulus with the MCMC \citep{Chauvenetcriterion1998PhT}. Chauvenet's criterion is widely used for outlier exclusion, and is commonly applied in various SN cosmological studies \citep[e.g.][]{chauvenet_snls_conley11,chauvenet_foley17or18,pahtheondataset,chauvent_vin23_desccsnmock}. Here we employ the Median Absolute Deviation ($\text{MAD}$) method to derive the median $\tilde{X}$ and the dispersion of the sample $\sigma_X=1.4826\,{\rm median}\left(\left|X_i-\tilde{X}\right|\right)$ for removing the outliers. The MAD can naturally suppress the weighting of the outliers, and could be more effective to remove the CCSN contamination. Based on Chauvenet's criterion, we remove all data points with $\left|X_i-\tilde{X}\right|>n\,\sigma_X$, where $n$ is found to be 3.68 for our sample size with total 2175 SNe. Besides, considering the error of the data point $E_{Xi}$, we will keep the data point when $E_{Xi}>\left|X_i-\tilde{X}\right|$.

Since $X_i=\mu_i$ here, as shown in Equation~(\ref{eq:tripp2}), it not only depends on the light curve parameters, but also the nuisance parameters $\alpha$, $\beta$ and $M_0$. These nuisance parameters should be determined or fitted along with the cosmological parameters in a cosmological model. We use the flat $w$CDM model for jointly fitting the cosmological and nuisance parameters, and deriving $\mu_i$ to further remove CCSNe. Note that the result of this removing process is not sensitive to a specific cosmological model, and we can obtain the same result by using the nonflat $\Lambda$CDM model.

We employed the MCMC method for parameter fitting using {\tt emcee} package \citep{emcee}. MCMC can derive the parameter probability distribution by illustrating the posterior probability $P(\boldsymbol{\theta} \mid\boldsymbol{D})$ for the parameter set $\boldsymbol{\theta}$ given the observational dataset $\boldsymbol{D}$, and we have
\begin{equation}
P(\boldsymbol{\theta}  \mid \boldsymbol{D})=\frac{\mathcal{L}(\boldsymbol{D} \mid \boldsymbol{\theta}) P(\boldsymbol{\theta})}{P(\boldsymbol{D})}.
\end{equation}
Here $P(\boldsymbol{\theta})$ is the parameter priors assumed to be uniform, $P(\boldsymbol{D})$ is the normalization factor which does not impact our analysis here, and \( \mathcal{L}(\boldsymbol{D} \mid \boldsymbol{\theta}) \sim e^{-\frac{1}{2}\chi^2} \) is the likelihood function, where $\chi^2$ is the chi-square given by
\begin{equation}
\begin{aligned}
& \chi^2=\sum_i^N\frac{\left[\mu_i-\mu_{\text{th}}(z_i)\right]^2}{\sigma_{i}^2}.
\end{aligned}
\end{equation}
The parameter ranges we set are $\Omega_{\rm M}\in(0,0.7)$, $\Omega_{\rm \Lambda}\in(0,2)$, $w\in(-2,0)$, $H_0\in(65,75)$, $M_0\in(-19.35,-19.15)$, and a range with $\pm5\%$ relative error for  $\alpha$ and $\beta$ in the flat $w$CDM and nonflat $\Lambda$CDM models, which are based on the current meansurements. We generate 80 chains, and each chain contains 10,000 chain points after burn-in. 

\begin{figure*}
    \centering
    \includegraphics[width=0.9\linewidth]{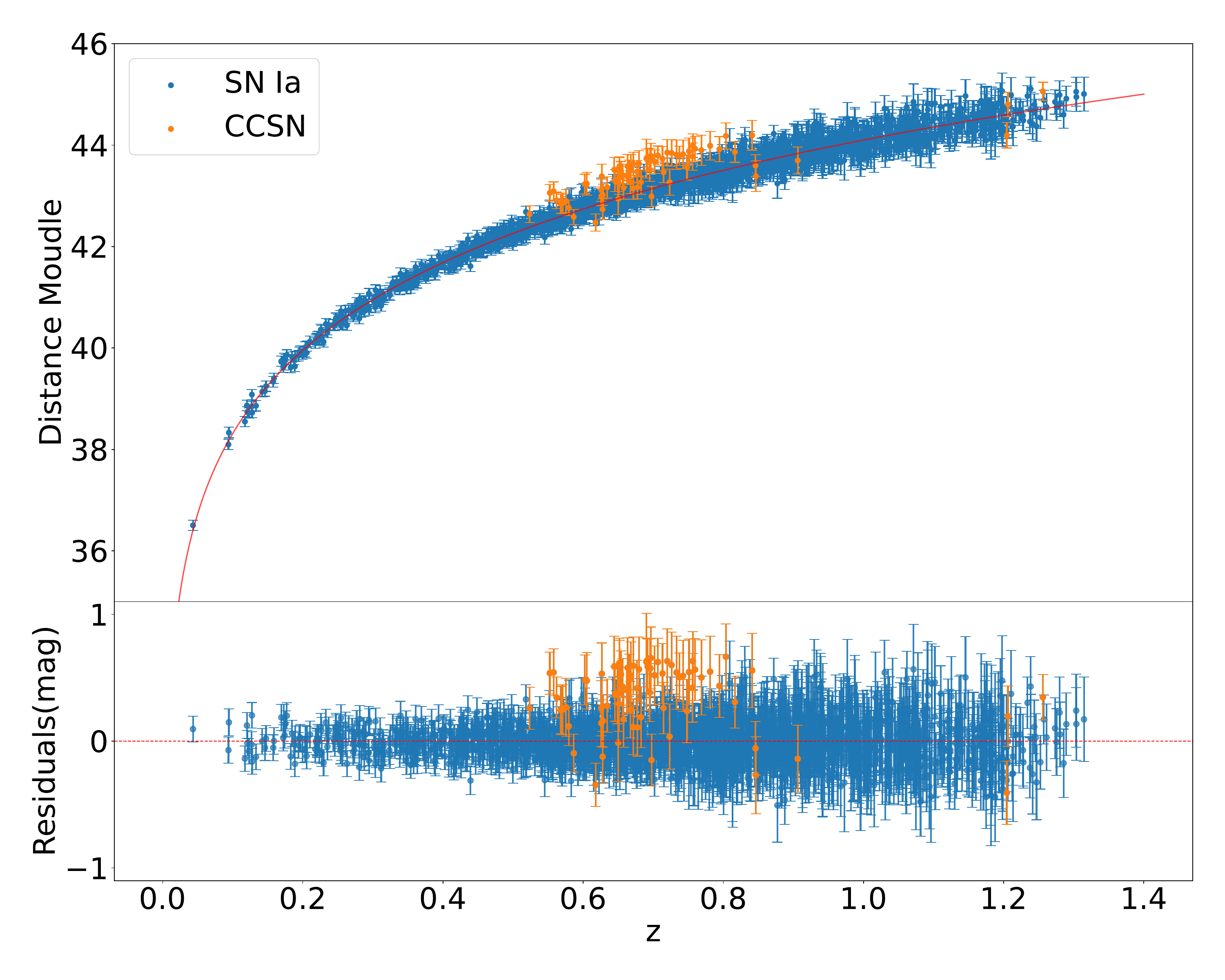}
    \caption{The Hubble diagram as a function of input redshifts for the SN sample, which is calibrated by the method based on Chauvenet's criterion. The blue and orange data points represent SNe Ia and CCSNe, respectively. The red curve is the theoretical result assuming the fiducial cosmology.}
    \label{fig:HDudf}
\end{figure*}

After obtaining $\mu_i$ by the MCMC process, we perform the calibration for further removing the CCSN contamination using the method based on Chauvenet's criterion. We find that 82 CCSNe and 8 SNe~Ia are removed from the dataset. Finally, there are 2085 SNe in the calibrated sample, and 73 CCSNe are left in the sample with the contamination fraction $\sim3.5\%$, compared to $\sim7.1\%$ before the calibration. We find that the SNe Ia at $z>0.5$ and $z>1$ can take the fractions of 83\% and 16\% of the total sample, respectively, which indicates that the CSST-UDF SN sample contains a large number of SNe Ia at high redshifts.

We show the Hubble diagram of the calibrated sample in Figure~\ref{fig:HDudf}. To compare with Figure~\ref{fig:hdall}, we use the fiducial values of the parameters to show the data points after calibration. We can find that the CCSN contamination is significantly reduced, and the remaining CCSN data are all close to the fiducial theoretical curve. The effectiveness of this method also can be indicated by the constraint results of the cosmological parameters as we discuss in the next section.


In addition, we also try other methods to further remove or reduce the CCSN contamination. For instance, we use the statistical method based on Bayesian estimation to reduce the contamination statistically on the cosmological parameter constraint  \citep{BEAMS,gong2010}. Additional $\chi^2$ with new parameters are added to include the effect of CCSN contamination in the fitting process, and the fraction of contamination also can be derived in this method. However, we find that this kind of method is not quite effective when multi types of CCSNe are considered, probably due to the complexity of the CCSN samples, and the method based on Chauvenet's criterion works better in this case. Besides, the machine learning method for supernova classification also can be used as an effective tool, given sufficient large and accurate training sets \citep[e.g.][]{PLAsTiCC2023}. We will leave it for future study.


\section{Cosmological constraint} \label{sec:result}

\begin{figure*}
  \centering
\begin{minipage}[b]{0.48\textwidth}  
 \includegraphics[width=1\linewidth]{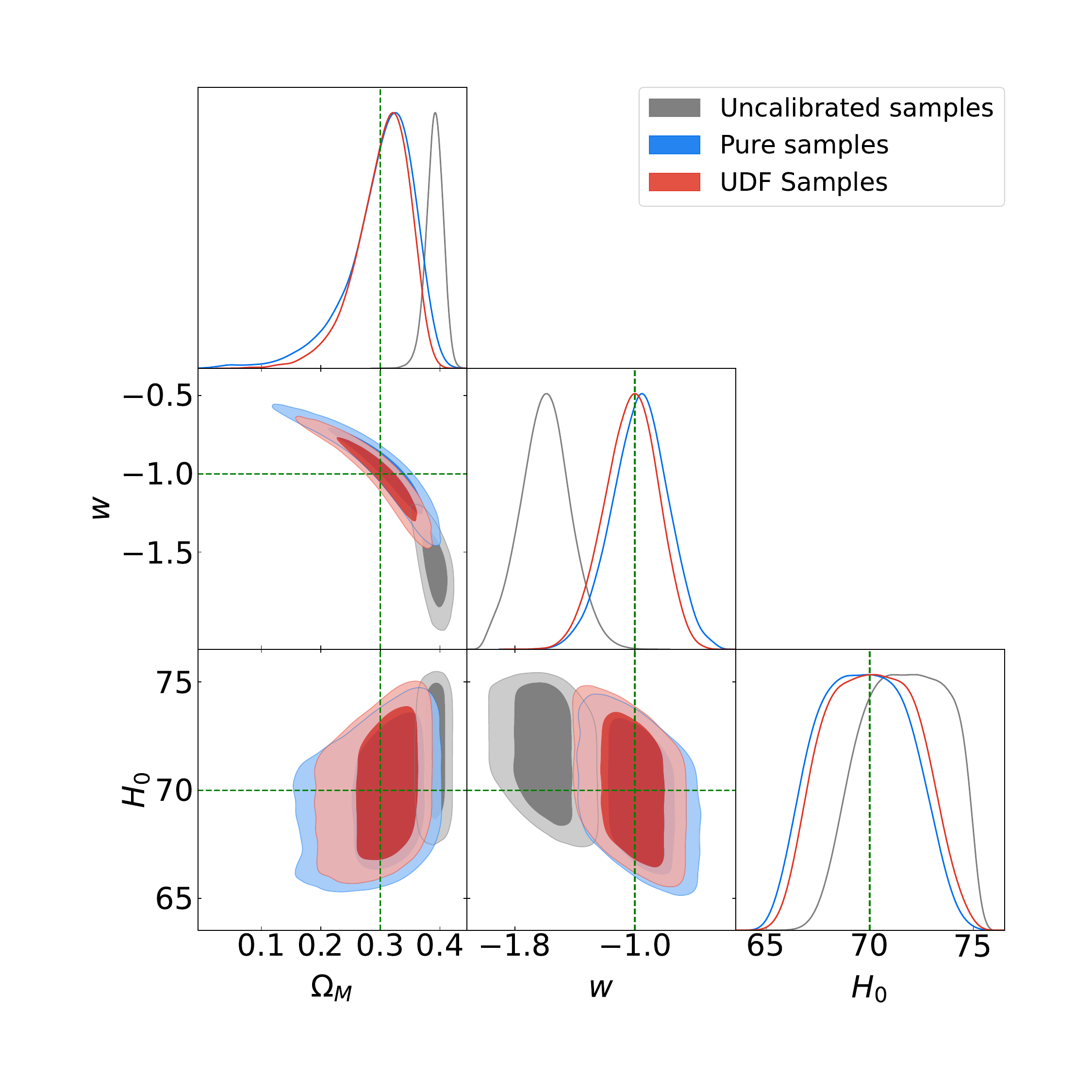}
\end{minipage}
\hfill
\begin{minipage}[b]{0.48\textwidth}
            \includegraphics[width=1\linewidth]{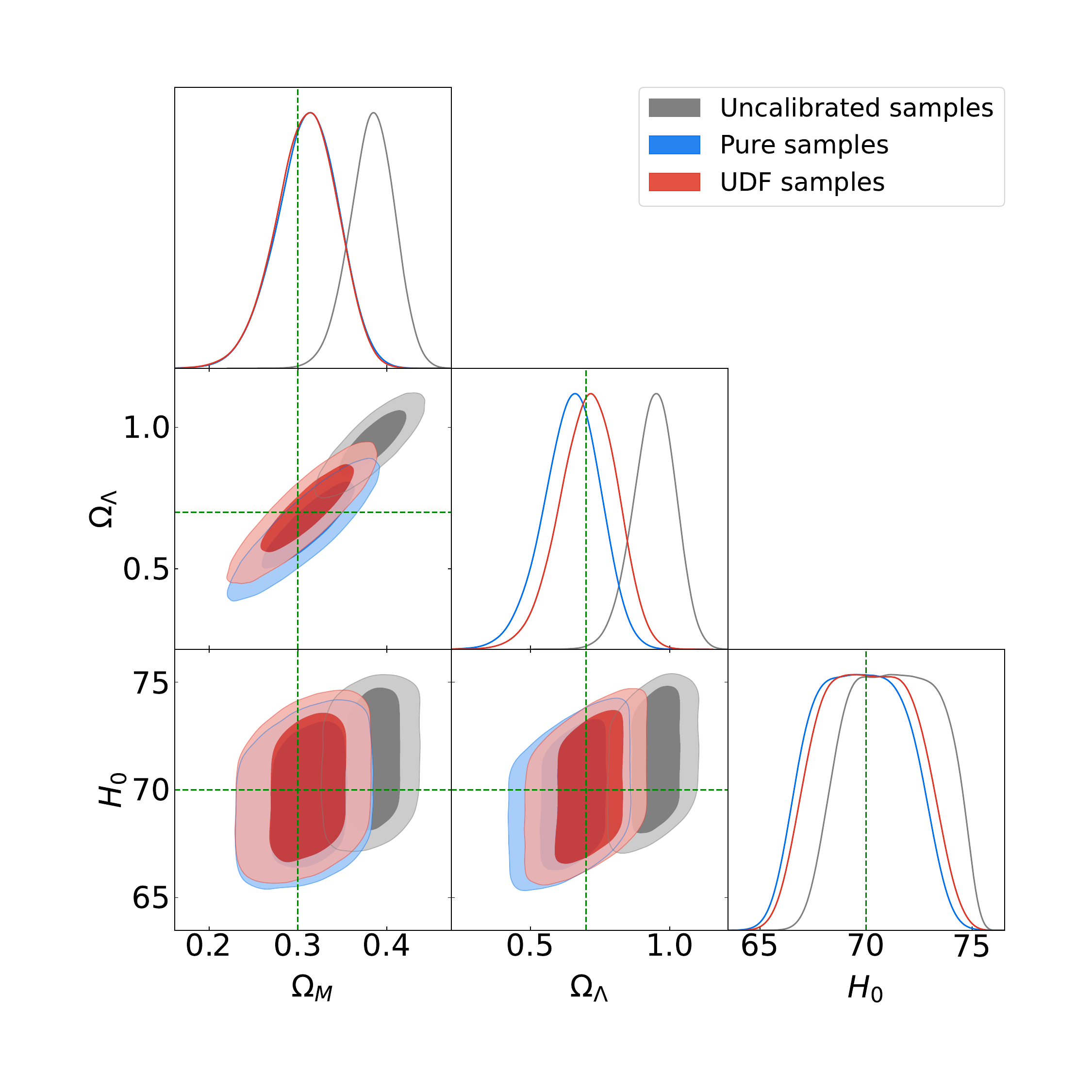}
\end{minipage}
  \caption{The predicted 1$\sigma$ and 2$\sigma$ contour maps and 1-D PDFs using the three different SN samples for the flat $w$CDM model (left panel) and nonflat $\Lambda$CDM model (right panel). The gray, blue, and red contours and curves represent the SN sample without calibration using the distance modulus data (i.e. uncalibrated sample), pure SN Ia sample, and calibrated UDF sample, respectively. The fiducial values of the cosmological parameters are shown in green lines for comparison.}  
\label{fig:3samples}
\end{figure*}

After obtaining the calibrated SN sample in the CSST-UDF survey, we investigate the constraints on the cosmological parameters.  In Figure~\ref{fig:3samples}, we present the constraint results for the flat $w$CDM (left panel) and nonflat $\Lambda$CDM (right panel) models using the pure SN~Ia sample (blue contours), uncalibrated SN sample (without calibration using the distance modulus data, in gray contours), and calibrated sample (i.e. UDF sample, in red contours). We can see that there are clear discrepancies for the constraint results of the uncalibrated SN sample from the fiducial values (green lines), which also indicates the necessity of the calibration after the light curve fitting process. On the other hand, the cosmological constraints using the pure SN Ia and calibrated UDF sample are in good agreement, and the contour maps and 1-D probability distribution functions (PDFs) are almost overlapped for some parameters. For the calibrated UDF sample, we find that the relative accuracies of the $1\sigma$ constraint on the cosmological parameters are $\sim14\%$, $18\%$ and $3.2\%$ for $\Omega_{\rm M}$, $w$ and $H_0$ in the flat $w$CDM model, and $\sim11\%$, $14\%$ and $3.2\%$ for $\Omega_{\rm M}$, $\Omega_{\Lambda}$ and $H_0$ in the nonflat $\Lambda$CDM model for the CSST-UDF SN survey. The details can be found in Table~\ref{tab:resut1}. Note that the constraint on $H_0$ could be worse if assuming a larger prior or range of $M_0$ in the fitting process, since there is a strong degeneracy between these two parameters. Compared to the current SN Ia surveys, e.g. Pantheon+ and DESSN-5YR \citep{pantheon22cosmosresult,descollaboration2024dark}, the CSST-UDF SN survey can improve the cosmological constraints by a factor of $\sim$2.

\begin{table}
    \centering
\caption{The best-fit values, 1$\sigma$ errors and relative accuracies for the relevant cosmological parameters in the flat $w$CDM and nonflat $\Lambda$CDM models using the SN Ia, BAO, and joint SN Ia+BAO data for the CSST-UDF surveys.}
\label{tab:resut1}
    \begin{tabular}{|c|c|c|c|} \hline 
         \multicolumn{4}{|c|}{Flat $w$CDM}\\\hline \hline 
         sample&  $\Omega_\text{m} $&  $w$&  $ H_\text{0}  $\\ \hline 
         SN only&  $0.309^{+0.036}_{-0.050}$ (14\%)&  $-1.02^{+0.17}_{-0.18}$ (18\%)&  $70.12^{2.25}_{-2.26}$ (3.2\%)\\ \hline 
         BAO only&  $0.290^{+0.031}_{-0.047} $ (13\%)&  $-1.03^{+0.28}_{-0.24} $ (25\%)&  $70.80^{+2.60}_{-2.84}$ (4.0\%)\\ \hline 
 SN+BAO& $0.301^{+0.026}_{-0.031} $ (9\%)& $-0.99^{+0.12}_{-0.13}$ (13\%)& $69.96^{+0.83}_{-0.83}$ (1.2\%)\\ \hline
 \multicolumn{4}{|c|}{Nonflat $\Lambda$CDM}\\\hline
 parameter
& $\Omega_\text{m} $
& $\Omega_\Lambda$
& $ H_\text{0}  $
\\\hline
 SN only
& $ 0.311^{+0.033}_{-0.035} $ (11\%)& $0.751^{+0.097}_{-0.105} $ (14\%)& $70.07^{+2.25}_{-2.23}$ (3.2\%)\\\hline
 BAO only
& $0.301^{+0.059}_{-0.059}$ (20\%)& 
$0.738^{+0.086}_{-0.090}$ (12\%)& $70.46^{+1.30}_{-1.30}$ (1.9\%)\\ \hline 
 SN+BAO& $0.290^{+0.026}_{-0.028}$ (9\%)& $0.691^{+0.064}_{-0.068 }$ (10\%)& $70.23^{+0.82}_{-0.82}$ (1.2\%)\\ \hline
    \end{tabular}
\end{table}

Besides, in order to explore the improvement of the constraint power, we also include the mock data of baryon acoustic oscillation (BAO) from the CSST spectroscopic wide-field galaxy survey, as another powerful measurement of the cosmic acceleration expansion history. Here we use the mock BAO data from \cite{miao2023forecasting}, and the angular diameter distance $d_{\rm A}$ and Hubble parameter $H(z)$ are derived at different redshift bins from $z=0$ to 1.5. The values of $d_{\rm A}$ and $H(z)$, and the relative errors are shown in Table~\ref{tab:BAODATA}. The expression of $d_{\rm A}$ is given by $d_{\rm A}=(1+z)^{-2}d_{\rm L}$. In the MCMC fitting process for the BAO data, we have $\chi_{\rm BAO}^2=\chi_{d_{\rm A}}^2+\chi_{H}^2$. More details of the generation of the CSST BAO mock data can be found in \cite{miao2023forecasting}. The BAO mock data we use are generated randomly from a Gaussian distributiaon based on the central values and errors shown in Table~\ref{tab:BAODATA}.

 \begin{table}
     \centering
\caption{The BAO data used in this study derived from \protect \cite{miao2023forecasting}. The units of $d_{\rm A}$  and $H_0$ are ${\rm Mpc}\,h^{-1}$ and km/s/Mpc, respectively.} 
\label{tab:BAODATA}
     \begin{tabular}{|c|c|c|c|c|} \hline  \hline 
          Redshift bin&  $ d_{\rm A} $&  $ d_{\rm A} $ error&  $H(z) $& $H(z) $ error\\ \hline 
          0 - 0.3&  539.2&  5.27\%&  75.3& 10.11\%\\ \hline 
          0.3 - 0.6&  1187.5&  2.1\%&  88.9& 4.28\%\\ \hline 
          0.6 - 0.9&  1513.7&  1.43\%&  1.6.3& 2.97\%\\ \hline 
          0.9 - 1.2&  1670.2&  1.23\%&  126.9& 2.54\%\\ \hline 
          1.2 - 1.5&  1734.4&  1.84\%&  150.0& 3.56\%\\ \hline
     \end{tabular}
 \end{table}

\begin{figure*}
  \centering
  \begin{minipage}[b]{0.48\textwidth}
    \centering
    \includegraphics[width=\textwidth]{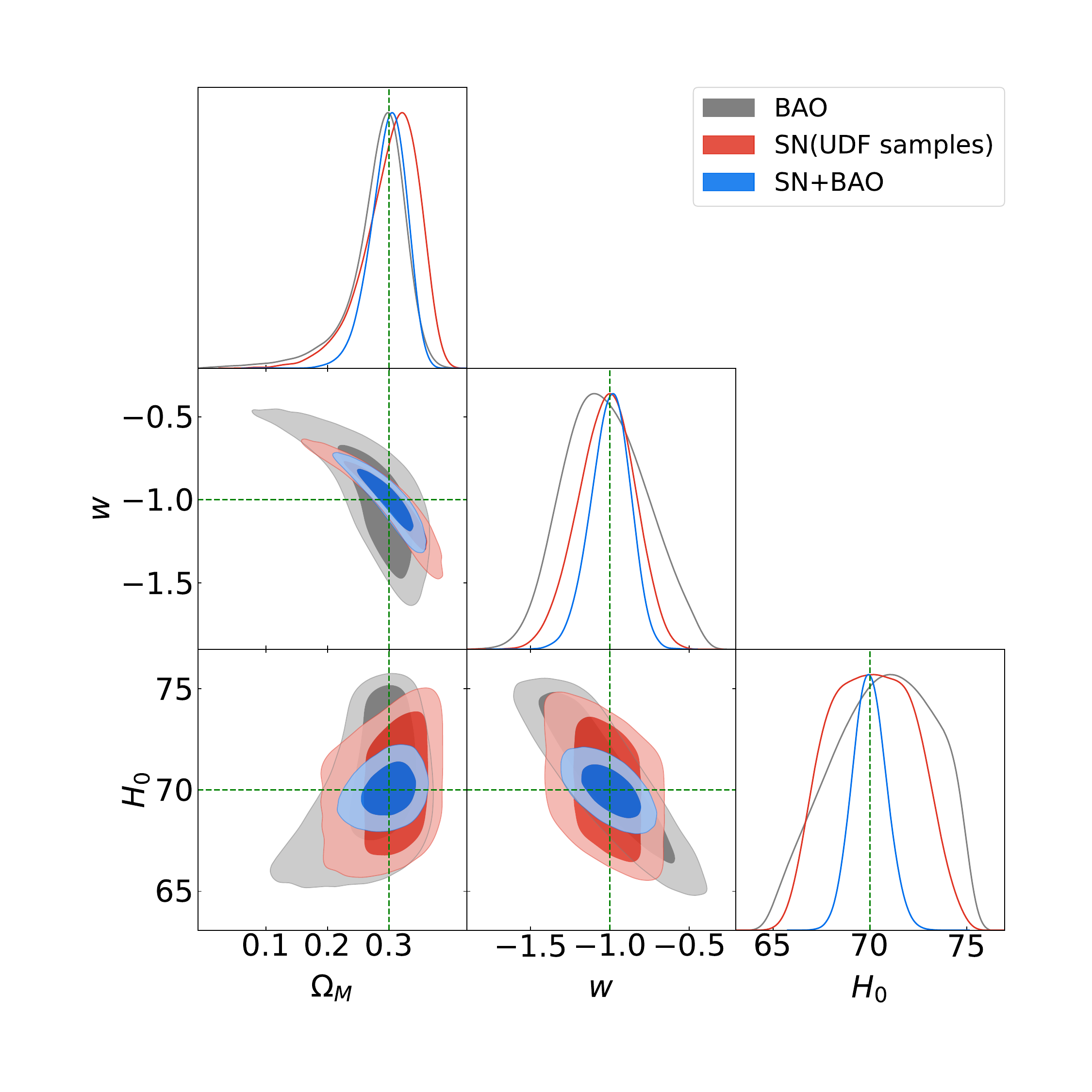}
  \end{minipage}
  \hfill
   \begin{minipage}[b]{0.48\textwidth}
  \includegraphics[width=\textwidth]{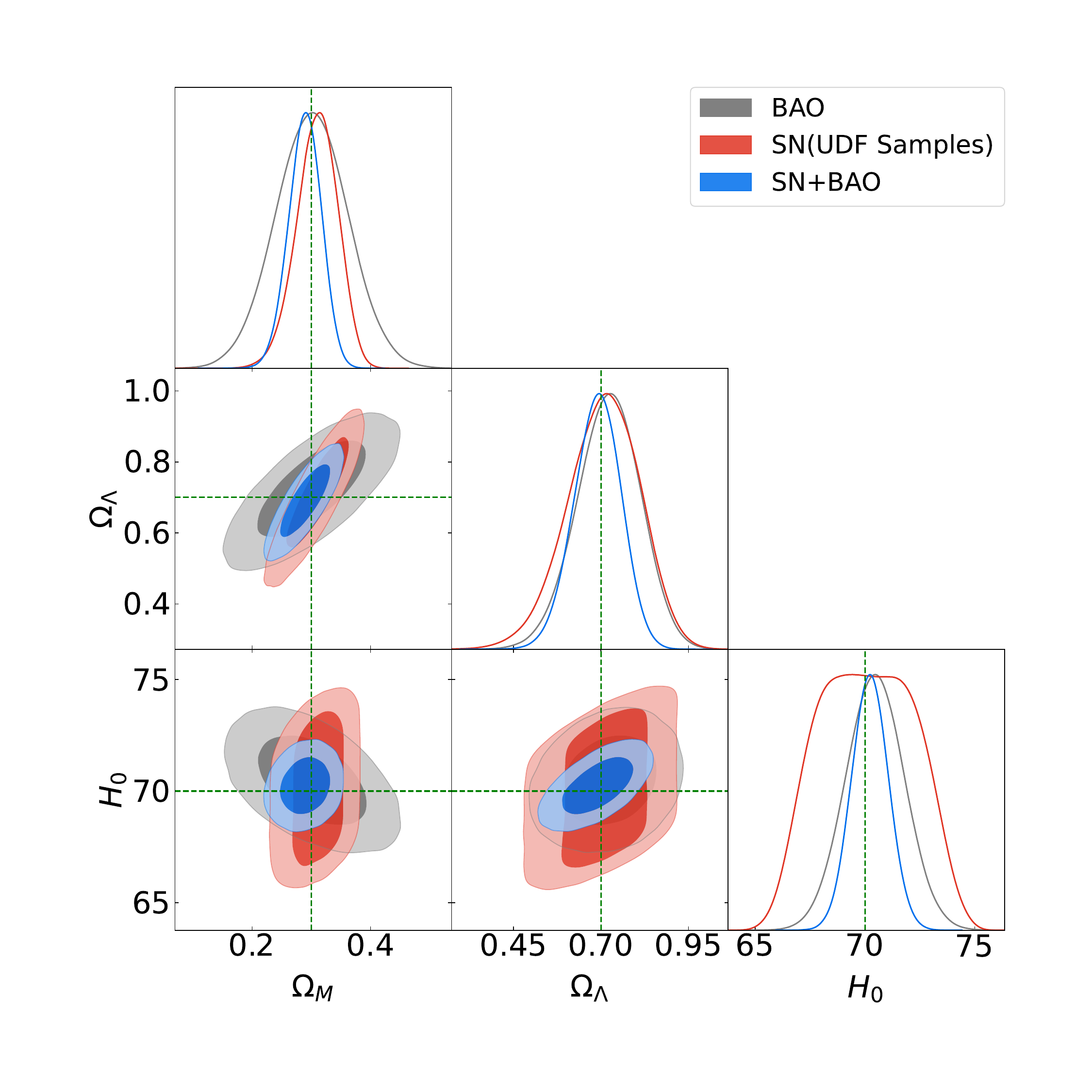}
  \end{minipage}
 \caption{\label{fig:csomosresult}The predicted 1$\sigma$ and 2$\sigma$ contour maps and 1-D PDFs for $\Omega_\text{M}$, $w$ and $H _\text{0} $ in the flat $w$CDM model (left panel) and $\Omega_\text{M} $, $\Omega_\Lambda$ and $H _\text{0}$ in the nonflat $\Lambda$CDM model (right panel) using the calibrated SN (red), BAO (gray), and joint (blue) survey data. The green dashed lines denote the fiducial values of the cosmological parameters for comparison.}
\end{figure*}

In Figure~\ref{fig:csomosresult}, the constraint results of the SN Ia, BAO, and joint data for the flat $w$CDM (left panel) and nonflat $\Lambda$CDM (right panel) models have been shown and compared. We find that the BAO data can be helpful to break the degeneracy between parameters, and make the constraints more stringent, especially for $H_0$. The constraint accuracies of the cosmological parameters after including the BAO data are $\sim9\%$, $13\%$ and $1.2\%$ for $\Omega_\text{M}$, $w$ and $H_0$ in the flat $w$CDM model, and $\sim9\%$, $10\%$ and $1.2\%$ for $\Omega_\text{M}$, $\Omega_{\Lambda}$ and $H_0$ in the nonflat $\Lambda$CDM model. Hence the constraints on the cosmological parameters can be improved by a factor of $\sim$1.4 for the SN Ia+BAO joint analysis, compared to the case with SN Ia data only.

\section{Summary and discussion}\label{sec:summary}

In this work, we explore the SN Ia observation and cosmological constraint for the CSST-UDF survey. We generate light curve mock data for both SNe Ia and CCSNe, based on the SED templates given in SALT3 and V19 models, the SN natural generation rates, luminosity functions, CSST instrumental design and CSST-UDF photometric survey strategy. Then high-quality light curve data are selected for the light curve analysis. By using the MCMC algorithm embedded in {\tt SNCosmo}, we fit the SN light curve data to derive the SN Ia light curve parameters (i.e. $z$, $t_0$, $m_B$, $x_1$ and $c$) and distinguish CCSNe as contamination. The SN samples that cannot match the SN~Ia model, and the values of $x_1$ and $c$ exceeding the SALT3 model requirements are identified and excluded as CCSNe. The resulting SN sample shows a $\sim$7.1\% contamination fraction.

To further reduce the contamination, we analyze the data of distance modulus, and adopt a method based on Chauvenet's criterion and considering the data error. After applying this method, we find that the contamination fraction can be effectively reduced to $\sim3.5\%$, resulting in a calibrated UDF sample with a SN Ia purity of $\sim$96.5\%. This calibrated CSST-UDF SN sample has a large number of SNe Ia at high redshifts with $\sim$16\% of SNe Ia are at $z>1$, which is expected to provide stringent constraints on cosmological parameter, such as dark energy equation of state.

Using this SN sample, the constraints on the cosmological parameters are investigated for the flat $w$CDM and nonflat $\Lambda$CDM models. We find that the constraint result of the calibrated sample is in good agreement with that derived from the pure SN~Ia sample. This indicates the effectiveness of our method for reducing the CCSN contamination. The constraint accuracy of the cosmological parameters is $\sim$15\% for $\Omega_{\rm M}$, $\Omega_{\Lambda}$, and $w$ in the CSST-UDF SN survey, which is better than the current SN Ia surveys by a factor of $\sim$2. When performing a joint analysis with the BAO data from the CSST spectroscopic galaxy survey, the accuracy can be improved by a factor of $\sim$1.4.

We note that there may be still room for improvement in the purity of the dataset. For example, the machine learning classifiers can be a good choice to further suppress the contamination. In the analysis of the DES data, i.e. DESSN-5YR \citep{descollaboration2024dark}, they can obtain a SN Ia purity above 98\% by using the machine learning classifier \citep{SuperNNova,chauvent_vin23_desccsnmock,DES5Year23}, which is better than our result ($\sim96.5\%$). In the future we will try the machine learning method to further improve the dataset quality.

Comparing to other Stage-IV surveys, e.g. LSST, the magnitude limit of the CSST-UDF single exposure is about one magnitude deeper than the LSST Deep Drilling Field (LSST-DDF) \citep{LSSTDDF2018,LSSTSN23}. This means that the sample with a larger high-$z$ SN Ia fraction can be obtained in the CSST-UDF survey. Besides, since the CSST-UDF survey has a great depth in the ultraviolet (UV) band with $NUV\sim25$ AB mag for a single exposure, the dust model can be well constrained, whcih would be quite helpful to calibrate this one of the biggest systematics in SN Ia cosmology.

We also notice that CSST can detect SNe Ia by the slitless spectroscopic observation, which is very helpful to identify SNe Ia and accurately measure the redshift. Although the magnitude limit of the CSST spectroscopic survey is lower than that in the photometric survey, most of SNe Ia at low redshifts could be measured by the spectroscopic survey. Hence, the current constraint results can be further improved by including the CSST spectroscopic data. Besides the CSST survey camera, CSST-MCI module also is capable to detect hundreds of SNe Ia, especially at high redshifts, which could effectively extend the CSST SN Ia sample. These imply that CSST has great potential in the SN Ia survey, and is expected to provide accurate measurements on cosmic distance and expansion history.

\section*{Acknowledgements}

MW and YG acknowledge the support from National Key R\&D Pro- gram of China grant Nos. 2022YFF0503404, 2020SKA0110402, and the CAS Project for Young Scientists in Basic Research (No. YSBR- 092). XC acknowledges the support of the National Natural Science Foundation of China through Grant Nos. 11473044 and 11973047, and the Chinese Academy of Science grants ZDKYYQ20200008, QYZDJ-SSW-SLH017, XDB 23040100, and XDA15020200. This work is also supported by science research grants from the China Manned Space Project with Grant Nos. CMS- CSST-2021-B01 and CMS-CSST-2021-A01.

\section*{Data Availability}

 The data that support the findings of this study are available from the corresponding author upon reasonable request.


\bibliographystyle{mnras}
\bibliography{A_cite} 

\begin{thebibliography}{}
\makeatletter
\relax
\def\mn@urlcharsother{\let\do\@makeother \do\$\do\&\do\#\do\^\do\_\do\%\do\~}
\def\mn@doi{\begingroup\mn@urlcharsother \@ifnextchar [ {\mn@doi@}
  {\mn@doi@[]}}
\def\mn@doi@[#1]#2{\def\@tempa{#1}\ifx\@tempa\@empty \href
  {http://dx.doi.org/#2} {doi:#2}\else \href {http://dx.doi.org/#2} {#1}\fi
  \endgroup}
\def\mn@eprint#1#2{\mn@eprint@#1:#2::\@nil}
\def\mn@eprint@arXiv#1{\href {http://arxiv.org/abs/#1} {{\tt arXiv:#1}}}
\def\mn@eprint@dblp#1{\href {http://dblp.uni-trier.de/rec/bibtex/#1.xml}
  {dblp:#1}}
\def\mn@eprint@#1:#2:#3:#4\@nil{\def\@tempa {#1}\def\@tempb {#2}\def\@tempc
  {#3}\ifx \@tempc \@empty \let \@tempc \@tempb \let \@tempb \@tempa \fi \ifx
  \@tempb \@empty \def\@tempb {arXiv}\fi \@ifundefined
  {mn@eprint@\@tempb}{\@tempb:\@tempc}{\expandafter \expandafter \csname
  mn@eprint@\@tempb\endcsname \expandafter{\@tempc}}}

\bibitem[\protect\citeauthoryear{{Abbott} et~al.,}{{Abbott}
  et~al.}{2019}]{DES_3YEARS_SN}
{Abbott} T.~M.~C.,  et~al., 2019, \mn@doi [\apjl] {10.3847/2041-8213/ab04fa},
  \href {https://ui.adsabs.harvard.edu/abs/2019ApJ...872L..30A} {872, L30}

\bibitem[\protect\citeauthoryear{{Akeson} et~al.,}{{Akeson}
  et~al.}{2019}]{WFIRST19}
{Akeson} R.,  et~al., 2019, \mn@doi [arXiv e-prints]
  {10.48550/arXiv.1902.05569}, \href
  {https://ui.adsabs.harvard.edu/abs/2019arXiv190205569A} {p. arXiv:1902.05569}

\bibitem[\protect\citeauthoryear{{Amendola} et~al.,}{{Amendola}
  et~al.}{2018}]{Euclid18}
{Amendola} L.,  et~al., 2018, \mn@doi [Living Reviews in Relativity]
  {10.1007/s41114-017-0010-3}, \href
  {https://ui.adsabs.harvard.edu/abs/2018LRR....21....2A} {21, 2}

\bibitem[\protect\citeauthoryear{{Barbary} et~al.,}{{Barbary}
  et~al.}{2016}]{SNCOSMO}
{Barbary} K.,  et~al., 2016, {SNCosmo: Python library for supernova cosmology},
  Astrophysics Source Code Library, record ascl:1611.017 (\mn@eprint {ascl}
  {1611.017})

\bibitem[\protect\citeauthoryear{{Bernstein} et~al.,}{{Bernstein}
  et~al.}{2012}]{DESHOSTGALAXY}
{Bernstein} J.~P.,  et~al., 2012, \mn@doi [\apj] {10.1088/0004-637X/753/2/152},
  \href {https://ui.adsabs.harvard.edu/abs/2012ApJ...753..152B} {753, 152}

\bibitem[\protect\citeauthoryear{{Betoule} et~al.,}{{Betoule}
  et~al.}{2014}]{JLA14}
{Betoule} M.,  et~al., 2014, \mn@doi [\aap] {10.1051/0004-6361/201423413},
  \href {https://ui.adsabs.harvard.edu/abs/2014A&A...568A..22B} {568, A22}

\bibitem[\protect\citeauthoryear{{Brout} \& {Scolnic}}{{Brout} \&
  {Scolnic}}{2021}]{BS21DUST}
{Brout} D.,  {Scolnic} D.,  2021, \mn@doi [\apj] {10.3847/1538-4357/abd69b},
  \href {https://ui.adsabs.harvard.edu/abs/2021ApJ...909...26B} {909, 26}

\bibitem[\protect\citeauthoryear{{Brout} et~al.,}{{Brout}
  et~al.}{2019}]{chauvenet_des_brout2019b}
{Brout} D.,  et~al., 2019, \mn@doi [\apj] {10.3847/1538-4357/ab08a0}, \href
  {https://ui.adsabs.harvard.edu/abs/2019ApJ...874..150B} {874, 150}

\bibitem[\protect\citeauthoryear{{Brout} et~al.,}{{Brout}
  et~al.}{2022}]{pantheon22cosmosresult}
{Brout} D.,  et~al., 2022, \mn@doi [\apj] {10.3847/1538-4357/ac8e04}, \href
  {https://ui.adsabs.harvard.edu/abs/2022ApJ...938..110B} {938, 110}

\bibitem[\protect\citeauthoryear{{Campbell} et~al.,}{{Campbell}
  et~al.}{2013}]{SDSS2_zerror}
{Campbell} H.,  et~al., 2013, \mn@doi [\apj] {10.1088/0004-637X/763/2/88},
  \href {https://ui.adsabs.harvard.edu/abs/2013ApJ...763...88C} {763, 88}

\bibitem[\protect\citeauthoryear{{Cao} et~al.,}{{Cao} et~al.}{2018}]{caoye18}
{Cao} Y.,  et~al., 2018, \mn@doi [\mnras] {10.1093/mnras/sty1980}, \href
  {https://ui.adsabs.harvard.edu/abs/2018MNRAS.480.2178C} {480, 2178}

\bibitem[\protect\citeauthoryear{{Cao}, {Gong}, {Liu}, {Cooray}, {Feng}  \&
  {Chen}}{{Cao} et~al.}{2022}]{CAOYE22}
{Cao} Y.,  {Gong} Y.,  {Liu} D.,  {Cooray} A.,  {Feng} C.,   {Chen} X.,  2022,
  \mn@doi [\mnras] {10.1093/mnras/stac151}, \href
  {https://ui.adsabs.harvard.edu/abs/2022MNRAS.511.1830C} {511, 1830}

\bibitem[\protect\citeauthoryear{{Conley} et~al.,}{{Conley}
  et~al.}{2011}]{chauvenet_snls_conley11}
{Conley} A.,  et~al., 2011, \mn@doi [\apjs] {10.1088/0067-0049/192/1/1}, \href
  {https://ui.adsabs.harvard.edu/abs/2011ApJS..192....1C} {192, 1}

\bibitem[\protect\citeauthoryear{{DES Collaboration} et~al.,}{{DES
  Collaboration} et~al.}{2024}]{descollaboration2024dark}
{DES Collaboration} et~al., 2024, \mn@doi [arXiv e-prints]
  {10.48550/arXiv.2401.02929}, \href
  {https://ui.adsabs.harvard.edu/abs/2024arXiv240102929D} {p. arXiv:2401.02929}

\bibitem[\protect\citeauthoryear{{Fitzpatrick}}{{Fitzpatrick}}{1999}]{f99}
{Fitzpatrick} E.~L.,  1999, \mn@doi [\pasp] {10.1086/316293}, \href
  {https://ui.adsabs.harvard.edu/abs/1999PASP..111...63F} {111, 63}

\bibitem[\protect\citeauthoryear{{Foley} et~al.,}{{Foley}
  et~al.}{2018}]{chauvenet_foley17or18}
{Foley} R.~J.,  et~al., 2018, \mn@doi [\mnras] {10.1093/mnras/stx3136}, \href
  {https://ui.adsabs.harvard.edu/abs/2018MNRAS.475..193F} {475, 193}

\bibitem[\protect\citeauthoryear{{Foreman-Mackey}, {Hogg}, {Lang}  \&
  {Goodman}}{{Foreman-Mackey} et~al.}{2013}]{emcee}
{Foreman-Mackey} D.,  {Hogg} D.~W.,  {Lang} D.,   {Goodman} J.,  2013, \mn@doi
  [PASP] {10.1086/670067}, 125, 306

\bibitem[\protect\citeauthoryear{{Gong}, {Cooray}  \& {Chen}}{{Gong}
  et~al.}{2010}]{gong2010}
{Gong} Y.,  {Cooray} A.,   {Chen} X.,  2010, \mn@doi [\apj]
  {10.1088/0004-637X/709/2/1420}, \href
  {https://ui.adsabs.harvard.edu/abs/2010ApJ...709.1420G} {709, 1420}

\bibitem[\protect\citeauthoryear{{Gong} et~al.,}{{Gong}
  et~al.}{2019}]{Gong2019}
{Gong} Y.,  et~al., 2019, \mn@doi [\apj] {10.3847/1538-4357/ab391e}, \href
  {https://ui.adsabs.harvard.edu/abs/2019ApJ...883..203G} {883, 203}

\bibitem[\protect\citeauthoryear{{Guy} et~al.,}{{Guy}
  et~al.}{2007}]{SALT2_2007}
{Guy} J.,  et~al., 2007, \mn@doi [\aap] {10.1051/0004-6361:20066930}, \href
  {https://ui.adsabs.harvard.edu/abs/2007A&A...466...11G} {466, 11}

\bibitem[\protect\citeauthoryear{{Guy} et~al.,}{{Guy} et~al.}{2010}]{SNLS2010}
{Guy} J.,  et~al., 2010, \mn@doi [\aap] {10.1051/0004-6361/201014468}, \href
  {https://ui.adsabs.harvard.edu/abs/2010A&A...523A...7G} {523, A7}

\bibitem[\protect\citeauthoryear{{Hicken} et~al.,}{{Hicken}
  et~al.}{2009}]{CFA3_2009Hicken}
{Hicken} M.,  et~al., 2009, \mn@doi [\apj] {10.1088/0004-637X/700/1/331}, \href
  {https://ui.adsabs.harvard.edu/abs/2009ApJ...700..331H} {700, 331}

\bibitem[\protect\citeauthoryear{{Hicken} et~al.,}{{Hicken}
  et~al.}{2012}]{CFA4_2012ApJS..200...12H}
{Hicken} M.,  et~al., 2012, \mn@doi [\apjs] {10.1088/0067-0049/200/2/12}, \href
  {https://ui.adsabs.harvard.edu/abs/2012ApJS..200...12H} {200, 12}

\bibitem[\protect\citeauthoryear{{Hlo{\v{z}}ek} et~al.,}{{Hlo{\v{z}}ek}
  et~al.}{2023}]{PLAsTiCC2023}
{Hlo{\v{z}}ek} R.,  et~al., 2023, \mn@doi [\apjs] {10.3847/1538-4365/accd6a},
  \href {https://ui.adsabs.harvard.edu/abs/2023ApJS..267...25H} {267, 25}

\bibitem[\protect\citeauthoryear{{Ivezi{\'c}} et~al.,}{{Ivezi{\'c}}
  et~al.}{2019}]{LSST2019}
{Ivezi{\'c}} {\v{Z}}.,  et~al., 2019, \mn@doi [\apj]
  {10.3847/1538-4357/ab042c}, \href
  {https://ui.adsabs.harvard.edu/abs/2019ApJ...873..111I} {873, 111}

\bibitem[\protect\citeauthoryear{{Jha} et~al.,}{{Jha}
  et~al.}{2006}]{CFA2_2006AJ....131..527J}
{Jha} S.,  et~al., 2006, \mn@doi [\aj] {10.1086/497989}, \href
  {https://ui.adsabs.harvard.edu/abs/2006AJ....131..527J} {131, 527}

\bibitem[\protect\citeauthoryear{{Jha}, {Riess}  \& {Kirshner}}{{Jha}
  et~al.}{2007}]{MLCS2K2}
{Jha} S.,  {Riess} A.~G.,   {Kirshner} R.~P.,  2007, \mn@doi [\apj]
  {10.1086/512054}, \href
  {https://ui.adsabs.harvard.edu/abs/2007ApJ...659..122J} {659, 122}

\bibitem[\protect\citeauthoryear{{Jones} et~al.,}{{Jones} et~al.}{2017}]{J17}
{Jones} D.~O.,  et~al., 2017, \mn@doi [\apj] {10.3847/1538-4357/aa767b}, \href
  {https://ui.adsabs.harvard.edu/abs/2017ApJ...843....6J} {843, 6}

\bibitem[\protect\citeauthoryear{{Kenworthy} et~al.,}{{Kenworthy}
  et~al.}{2021}]{SALT3_Kenworthy2021}
{Kenworthy} W.~D.,  et~al., 2021, \mn@doi [\apj] {10.3847/1538-4357/ac30d8},
  \href {https://ui.adsabs.harvard.edu/abs/2021ApJ...923..265K} {923, 265}

\bibitem[\protect\citeauthoryear{{Kessler} et~al.,}{{Kessler}
  et~al.}{2009}]{SNANA}
{Kessler} R.,  et~al., 2009, \mn@doi [\pasp] {10.1086/605984}, \href
  {https://ui.adsabs.harvard.edu/abs/2009PASP..121.1028K} {121, 1028}

\bibitem[\protect\citeauthoryear{Kunz, Bassett  \& Hlozek}{Kunz
  et~al.}{2007}]{BEAMS}
Kunz M.,  Bassett B.~A.,   Hlozek R.~A.,  2007, Physical Review D, 75, 103508

\bibitem[\protect\citeauthoryear{{Laureijs} et~al.,}{{Laureijs}
  et~al.}{2011}]{Euclid11}
{Laureijs} R.,  et~al., 2011, \mn@doi [arXiv e-prints]
  {10.48550/arXiv.1110.3193}, \href
  {https://ui.adsabs.harvard.edu/abs/2011arXiv1110.3193L} {p. arXiv:1110.3193}

\bibitem[\protect\citeauthoryear{{L{\'e}get} et~al.,}{{L{\'e}get}
  et~al.}{2020}]{SUGAR20}
{L{\'e}get} P.~F.,  et~al., 2020, \mn@doi [\aap] {10.1051/0004-6361/201834954},
  \href {https://ui.adsabs.harvard.edu/abs/2020A&A...636A..46L} {636, A46}

\bibitem[\protect\citeauthoryear{{Li}, {Li}, {Zhang}, {Vink{\'o}},
  {Reg{\H{o}}s}, {Wang}, {Xi}  \& {Zhan}}{{Li} et~al.}{2023}]{LISHIYU}
{Li} S.-Y.,  {Li} Y.-L.,  {Zhang} T.,  {Vink{\'o}} J.,  {Reg{\H{o}}s} E.,
  {Wang} X.,  {Xi} G.,   {Zhan} H.,  2023, \mn@doi [Science China Physics,
  Mechanics, and Astronomy] {10.1007/s11433-022-2018-0}, \href
  {https://ui.adsabs.harvard.edu/abs/2023SCPMA..6629511L} {66, 229511}

\bibitem[\protect\citeauthoryear{{Lochner} et~al.,}{{Lochner}
  et~al.}{2018}]{LSSTDDF2018}
{Lochner} M.,  et~al., 2018, \mn@doi [arXiv e-prints]
  {10.48550/arXiv.1812.00515}, \href
  {https://ui.adsabs.harvard.edu/abs/2018arXiv181200515L} {p. arXiv:1812.00515}

\bibitem[\protect\citeauthoryear{{Lucas} \& {Ryon}}{{Lucas} \&
  {Ryon}}{2022}]{ACSBOOK}
{Lucas} R.~A.,  {Ryon} J.~E.,  2022, in , Vol.~11, ACS Data Handbook v. 11.0.
p.~11

\bibitem[\protect\citeauthoryear{Miao, Gong, Chen, Huang, Li  \& Zhan}{Miao
  et~al.}{2023}]{miao2023forecasting}
Miao H.,  Gong Y.,  Chen X.,  Huang Z.,  Li X.-D.,   Zhan H.,  2023,
  Forecasting the BAO Measurements of the CSST galaxy and AGN Spectroscopic
  Surveys (\mn@eprint {arXiv} {2311.16903})

\bibitem[\protect\citeauthoryear{{Miknaitis} et~al.,}{{Miknaitis}
  et~al.}{2007}]{ESSENCE_2007ApJ...666..674M}
{Miknaitis} G.,  et~al., 2007, \mn@doi [\apj] {10.1086/519986}, \href
  {https://ui.adsabs.harvard.edu/abs/2007ApJ...666..674M} {666, 674}

\bibitem[\protect\citeauthoryear{{Mitra}, {Kessler}, {More}, {Hlozek}  \& {LSST
  Dark Energy Science Collaboration}}{{Mitra} et~al.}{2023}]{LSSTSN23}
{Mitra} A.,  {Kessler} R.,  {More} S.,  {Hlozek} R.,   {LSST Dark Energy
  Science Collaboration} 2023, \mn@doi [\apj] {10.3847/1538-4357/acb057}, \href
  {https://ui.adsabs.harvard.edu/abs/2023ApJ...944..212M} {944, 212}

\bibitem[\protect\citeauthoryear{{M{\"o}ller} \& {de
  Boissi{\`e}re}}{{M{\"o}ller} \& {de Boissi{\`e}re}}{2020}]{SuperNNova}
{M{\"o}ller} A.,  {de Boissi{\`e}re} T.,  2020, \mn@doi [\mnras]
  {10.1093/mnras/stz3312}, \href
  {https://ui.adsabs.harvard.edu/abs/2020MNRAS.491.4277M} {491, 4277}

\bibitem[\protect\citeauthoryear{{M{\"o}ller} et~al.,}{{M{\"o}ller}
  et~al.}{2022}]{DES5Year23}
{M{\"o}ller} A.,  et~al., 2022, \mn@doi [\mnras] {10.1093/mnras/stac1691},
  \href {https://ui.adsabs.harvard.edu/abs/2022MNRAS.514.5159M} {514, 5159}

\bibitem[\protect\citeauthoryear{{Perlmutter} et~al.,}{{Perlmutter}
  et~al.}{1999}]{Perlmutter1999}
{Perlmutter} S.,  et~al., 1999, \mn@doi [\apj] {10.1086/307221}, \href
  {https://ui.adsabs.harvard.edu/abs/1999ApJ...517..565P} {517, 565}

\bibitem[\protect\citeauthoryear{{Popovic}, {Brout}, {Kessler}, {Scolnic}  \&
  {Lu}}{{Popovic} et~al.}{2021}]{Ppo21DUST}
{Popovic} B.,  {Brout} D.,  {Kessler} R.,  {Scolnic} D.,   {Lu} L.,  2021,
  \mn@doi [\apj] {10.3847/1538-4357/abf14f}, \href
  {https://ui.adsabs.harvard.edu/abs/2021ApJ...913...49P} {913, 49}

\bibitem[\protect\citeauthoryear{{Popovic}, {Brout}, {Kessler}  \&
  {Scolnic}}{{Popovic} et~al.}{2023}]{pop23dust}
{Popovic} B.,  {Brout} D.,  {Kessler} R.,   {Scolnic} D.,  2023, \mn@doi [\apj]
  {10.3847/1538-4357/aca273}, \href
  {https://ui.adsabs.harvard.edu/abs/2023ApJ...945...84P} {945, 84}

\bibitem[\protect\citeauthoryear{{Rest} et~al.,}{{Rest}
  et~al.}{2014}]{PANSTAR1_2014ApJ...795...44R}
{Rest} A.,  et~al., 2014, \mn@doi [\apj] {10.1088/0004-637X/795/1/44}, \href
  {https://ui.adsabs.harvard.edu/abs/2014ApJ...795...44R} {795, 44}

\bibitem[\protect\citeauthoryear{{Riess} et~al.,}{{Riess}
  et~al.}{1998}]{Riess98}
{Riess} A.~G.,  et~al., 1998, \mn@doi [\aj] {10.1086/300499}, \href
  {https://ui.adsabs.harvard.edu/abs/1998AJ....116.1009R} {116, 1009}

\bibitem[\protect\citeauthoryear{{Riess} et~al.,}{{Riess}
  et~al.}{1999}]{CFA1_RIESS1999AJ....117..707R}
{Riess} A.~G.,  et~al., 1999, \mn@doi [\aj] {10.1086/300738}, \href
  {https://ui.adsabs.harvard.edu/abs/1999AJ....117..707R} {117, 707}

\bibitem[\protect\citeauthoryear{{Riess} et~al.,}{{Riess}
  et~al.}{2004}]{goods2004}
{Riess} A.~G.,  et~al., 2004, \mn@doi [\apj] {10.1086/383612}, \href
  {https://ui.adsabs.harvard.edu/abs/2004ApJ...607..665R} {607, 665}

\bibitem[\protect\citeauthoryear{{Riess} et~al.,}{{Riess}
  et~al.}{2007}]{GOODS_Riess17...659...98R}
{Riess} A.~G.,  et~al., 2007, \mn@doi [\apj] {10.1086/510378}, \href
  {https://ui.adsabs.harvard.edu/abs/2007ApJ...659...98R} {659, 98}

\bibitem[\protect\citeauthoryear{{Riess} et~al.,}{{Riess}
  et~al.}{2018}]{Candels_Riess2018ApJ...853..126R}
{Riess} A.~G.,  et~al., 2018, \mn@doi [\apj] {10.3847/1538-4357/aaa5a9}, \href
  {https://ui.adsabs.harvard.edu/abs/2018ApJ...853..126R} {853, 126}

\bibitem[\protect\citeauthoryear{{Riess}, {Casertano}, {Yuan}, {Macri}  \&
  {Scolnic}}{{Riess} et~al.}{2019}]{SH0ES_2019ApJ...876...85R}
{Riess} A.~G.,  {Casertano} S.,  {Yuan} W.,  {Macri} L.~M.,   {Scolnic} D.,
  2019, \mn@doi [\apj] {10.3847/1538-4357/ab1422}, \href
  {https://ui.adsabs.harvard.edu/abs/2019ApJ...876...85R} {876, 85}

\bibitem[\protect\citeauthoryear{{Riess} et~al.,}{{Riess}
  et~al.}{2022}]{MABS-19.253RIESS}
{Riess} A.~G.,  et~al., 2022, \mn@doi [\apjl] {10.3847/2041-8213/ac5c5b}, \href
  {https://ui.adsabs.harvard.edu/abs/2022ApJ...934L...7R} {934, L7}

\bibitem[\protect\citeauthoryear{{Rodney} et~al.,}{{Rodney}
  et~al.}{2014}]{SN_RATE14}
{Rodney} S.~A.,  et~al., 2014, \mn@doi [\aj] {10.1088/0004-6256/148/1/13},
  \href {https://ui.adsabs.harvard.edu/abs/2014AJ....148...13R} {148, 13}

\bibitem[\protect\citeauthoryear{{Scolnic} et~al.,}{{Scolnic}
  et~al.}{2022}]{pahtheondataset}
{Scolnic} D.,  et~al., 2022, \mn@doi [\apj] {10.3847/1538-4357/ac8b7a}, \href
  {https://ui.adsabs.harvard.edu/abs/2022ApJ...938..113S} {938, 113}

\bibitem[\protect\citeauthoryear{{Strolger} et~al.,}{{Strolger}
  et~al.}{2015}]{CCRATE——strolger15}
{Strolger} L.-G.,  et~al., 2015, \mn@doi [\apj] {10.1088/0004-637X/813/2/93},
  \href {https://ui.adsabs.harvard.edu/abs/2015ApJ...813...93S} {813, 93}

\bibitem[\protect\citeauthoryear{{Suzuki} et~al.,}{{Suzuki}
  et~al.}{2012}]{SCP_suzuki_2012ApJ...746...85S}
{Suzuki} N.,  et~al., 2012, \mn@doi [\apj] {10.1088/0004-637X/746/1/85}, \href
  {https://ui.adsabs.harvard.edu/abs/2012ApJ...746...85S} {746, 85}

\bibitem[\protect\citeauthoryear{{The LSST Dark Energy Science Collaboration}
  et~al.,}{{The LSST Dark Energy Science Collaboration}
  et~al.}{2018}]{LSSTDESCRequirement}
{The LSST Dark Energy Science Collaboration} et~al., 2018, \mn@doi [arXiv
  e-prints] {10.48550/arXiv.1809.01669}, \href
  {https://ui.adsabs.harvard.edu/abs/2018arXiv180901669T} {p. arXiv:1809.01669}

\bibitem[\protect\citeauthoryear{{Thompson}}{{Thompson}}{1998}]{Chauvenetcriterion1998PhT}
{Thompson} W.,  1998, \mn@doi [Physics Today] {10.1063/1.882103}, \href
  {https://ui.adsabs.harvard.edu/abs/1998PhT....51a..57T} {51, 57}

\bibitem[\protect\citeauthoryear{{Vincenzi}, {Sullivan}, {Firth},
  {Guti{\'e}rrez}, {Frohmaier}, {Smith}, {Angus}  \& {Nichol}}{{Vincenzi}
  et~al.}{2019}]{V19}
{Vincenzi} M.,  {Sullivan} M.,  {Firth} R.~E.,  {Guti{\'e}rrez} C.~P.,
  {Frohmaier} C.,  {Smith} M.,  {Angus} C.,   {Nichol} R.~C.,  2019, \mn@doi
  [\mnras] {10.1093/mnras/stz2448}, \href
  {https://ui.adsabs.harvard.edu/abs/2019MNRAS.489.5802V} {489, 5802}

\bibitem[\protect\citeauthoryear{{Vincenzi} et~al.,}{{Vincenzi}
  et~al.}{2023}]{chauvent_vin23_desccsnmock}
{Vincenzi} M.,  et~al., 2023, \mn@doi [\mnras] {10.1093/mnras/stac1404}, \href
  {https://ui.adsabs.harvard.edu/abs/2023MNRAS.518.1106V} {518, 1106}

\bibitem[\protect\citeauthoryear{Vincenzi et~al.,}{Vincenzi
  et~al.}{2024}]{vincenzi2024desSystematic}
Vincenzi M.,  et~al., 2024, The Dark Energy Survey Supernova Program:
  Cosmological Analysis and Systematic Uncertainties (\mn@eprint {arXiv}
  {2401.02945})

\bibitem[\protect\citeauthoryear{{Zhan}}{{Zhan}}{2011}]{huzhan2011}
{Zhan} H.,  2011, \mn@doi [Scientia Sinica Physica, Mechanica \& Astronomica]
  {10.1360/132011-961}, \href
  {https://ui.adsabs.harvard.edu/abs/2011SSPMA..41.1441Z} {41, 1441}

\bibitem[\protect\citeauthoryear{Zhan}{Zhan}{2021}]{HuZhan2021}
Zhan H.,  2021, \mn@doi [Chinese Science Bulletin]
  {https://doi.org/10.1360/TB-2021-0016}, 66, 1290

\makeatother
\end{thebibliography}



\bsp	
\label{lastpage}
\end{document}